\documentclass[11pt,a4paper]{article}

\usepackage{hyperref}

\usepackage{vmargin}
\setmarginsrb{2.6cm}{1.7cm}{2.6cm}{1.7cm}{1cm}{1cm}{1cm}{1.6cm}

\usepackage{mathrsfs}   
\usepackage{dsfont}
\usepackage{euscript}
\usepackage{yfonts}

\usepackage[fleqn]{amsmath}
\usepackage{amsthm,amssymb,epsfig,latexsym}

\newtheorem{prop}{Proposition}[section]
\newtheorem{lemma}{Lemma}[section]

{\theoremstyle{remark}
\newtheorem{rem}{Remark}[section]}
\def\Proof{\medskip\noindent {\sl Proof --- \ }}
\def\qed{\hfill\nobreak\hbox{$\square$}\par\medbreak}

 \makeatletter
 \@addtoreset{equation}{section}
 \makeatother

\newcommand{\num}{\\\rule{0pt}{20pt}}

\newcommand{\dis}{\displaystyle}
\newcommand{\eq}[1]{(\ref{#1})}
\newcommand{\tr}{\mathop{\rm tr}}

\newcommand{\bra}[1]{\langle\,#1\,|}
\newcommand{\ket}[1]{|\,#1\,\rangle}
\def\ba{\begin{array}}
\def\ea{\end{array}}

\def\det{\operatorname{det}}

\newcommand{\f}[2]{{\ensuremath{%
    \mathchoice%
    {\dfrac{#1}{#2}}
    {\dfrac{#1}{#2}}
    {\frac{#1}{#2}}
    {\frac{#1}{#2}}
}}}

\newcommand{\tf}[2]{\ensuremath{#1/#2}}
\newcommand{\pa}[1]{\ensuremath{\left(#1\right)}}
\newcommand{\paa}[1]{\ensuremath{\left\{#1\right\}}}
\newcommand{\pac}[1]{\ensuremath{\left[#1\right]}}
\newcommand{\paf}[2]{\ensuremath{\left(\f{#1}{#2}\right)}}

\def\be{\beta}

\def\Ga{\Gamma}

\def\De{\Delta}
\def\eps{\epsilon}
\def\veps{\varepsilon}
\def\la{\lambda}

\def\sg{\sigma}

\def\th{\theta}

\newcommand{\wt}[1]{\ensuremath{\widetilde{#1}}}

\newcommand{\Int}[2]{\ensuremath{\int\limits_{#1}^{#2}}}

\newcommand{\sul}[2]{\ensuremath{\sum\limits_{#1}^{#2}}}

\newcommand{\R}{\ensuremath{\mathbb{R}}}

\let\tend=\rightarrow

\def\tr{\operatorname{tr}}

\newcommand{\abs}[1]{\ensuremath{\left| #1 \right|}}
\newcommand{\norm}[1]{\ensuremath{\left\|#1\right\|}}

\newcommand{\dd}{\mathrm{d}}
\newcommand{\e}[1]{\ensuremath{\mathrm{#1}}}

\begin{document}
\begin{flushright}
LPENSL-TH-03/09\\
\end{flushright}
\par \vskip .1in \noindent

\vspace{24pt}

\begin{center}
\begin{LARGE}
\vspace*{1cm}
 {On the thermodynamic limit of
 form factors in the massless XXZ Heisenberg chain}
\end{LARGE}

\vspace{50pt}

\begin{large}

{\bf N.~Kitanine}\footnote[1]{LPTM, UMR 8089, CNRS et Universit\'e
de Cergy-Pontoise, France, kitanine@u-cergy.fr},~~
{\bf K.~K.~Kozlowski}\footnote[2]{ Laboratoire de Physique, UMR 5672, CNRS et  ENS
Lyon, France,
 karol.kozlowski@ens-lyon.fr},~~
{\bf J.~M.~Maillet}\footnote[3]{ Laboratoire de Physique, UMR 5672, CNRS et ENS Lyon,  France,
 maillet@ens-lyon.fr},\\
{\bf N.~A.~Slavnov}\footnote[4]{ Steklov Mathematical Institute,
Moscow, Russia, nslavnov@mi.ras.ru},~~
{\bf V.~Terras}\footnote[5]{ Laboratoire de Physique, UMR 5672, 
CNRS et ENS Lyon,  France, veronique.terras@ens-lyon.fr, on leave of
absence from LPTA, UMR 5207, CNRS et Universit\'e Montpellier II}
\par

\end{large}

\vspace{80pt}

\centerline{\bf Abstract} \vspace{1cm}
\parbox{12cm}{\small We consider the problem of computing form factors of the massless XXZ Heisenberg spin-1/2  chain in a magnetic field in the (thermodynamic) limit where the size $M$ of the chain becomes large. For that purpose, we take the particular example of  the
matrix element  of the operator $\sigma^z$  between the ground state and an excited
state with one particle and one hole located at the opposite ends of
the Fermi interval (umklapp-type term). We  exhibit its power-law decrease in terms of the size of the chain $M$, and compute the corresponding exponent and amplitude. As a consequence, we show that this form factor is directly related to the amplitude of the leading oscillating term in the long-distance asymptotic
expansion of the correlation function $\langle \sigma_1^z
\sigma_{m+1}^z \rangle$. }
\end{center}

\newpage

\section{Introduction}

The main purpose of this article is to open a way for the study of the form factors (or in general of the matrix elements of local operators) of the massless XXZ Heisenberg chain in a magnetic field in the limit where the size of the chain becomes large. For that purpose we will consider a particular matrix element of the local spin operator $\sigma^z$ between the ground state and an excited
state with one particle and one hole located at the opposite ends of
the Fermi interval (umklapp-type term).
Although the present article is devoted to this special case, it will be clear that the method we propose can be applied to more general cases as well. 

As already mentioned in our recent work
\cite{KitKMST08b}, this particular form factor is of direct interest for the computation of the asymptotic behavior of the $\sigma^z$ two-point correlation function.
Let us briefly recall how it appears in this context.
In \cite{KitKMST08b} we explained how to compute, from first
principles, the long-distance asymptotic behavior of some two-point
correlation functions: the longitudinal spin-spin correlation
function in the massless phase of the XXZ spin-$\tf{1}{2}$
Heisenberg chain and also the density-density correlation function of
the quantum one-dimensional Bose gas.
In particular, for the spin chain, we obtained that the correlation function of the third
components of spin behaves at large distance $m$ as
 \begin{equation}\label{9-corr-funct}
 \langle\sigma_1^z\sigma_{m+1}^z\rangle-\langle\sigma_1^z\rangle^2=
 -\frac{2{\cal Z}^2}{\pi^2 m^2}+2|F_\sigma|^2\cdot
 \frac{\cos(2mp_{{}_F}) }{m^{2{\cal Z}^2}} \; +  \; \text{corrections}.
 \end{equation}
In this formula $p_{{}_F}$ stands for the Fermi momentum and ${\cal Z}$ represents the value of the
dressed charge on the Fermi boundary, the precise definition of these
quantities being postponed to Section~\ref{sec-XXZ-lim}.
We announced in \cite{KitKMST08b}
that the coefficient $F_\sigma$ in \eqref{9-corr-funct} is related to the thermodynamic limit
of the properly normalized
aforementioned special form factor of the $\sigma^z$ operator.
We prove this statement in this article.

The fact that the coefficient $F_\sigma$ should be related to this
form factor is not really surprising. It is well known that the
leading asymptotic behavior of the correlation functions is defined
by low-energy excitations in the spectrum of the Hamiltonian.
From this assumption, the form of the asymptotics \eqref{9-corr-funct} was
predicted by the Luttinger liquid approach
\cite{LutP75,Hal80,Hal81a,Hal81b,BogIK86,BogIR87,IzeKR89} and the
conformal field theory \cite{Car84,Aff85,Car86,BloCN86} together
with the analysis of finite size corrections
\cite{DeVW85,Woy87,WoyE87, DesDV88,WoyET89, KluB90,KluBP91,
KluWZ93,DesDV95}. There the oscillating term in the asymptotic
formula corresponds to the process of  a particle jumping from one
Fermi boundary to the other. One can expect therefore that the numerical
coefficient in this term should be somehow related to the
corresponding form factor.

However, the precise relationship between the coefficient $F_\sigma$
and the form factor under consideration  has been missing up to now.
The first reason is that  the methods mentioned above hardly give any
prediction for the exact value of the coefficient $F_\sigma$: 
an expression for this amplitude was already given in  \cite{Luk99,LukT03}, but only
for the case of zero magnetic field, and its physical meaning was not clear.
The second reason is the absence of sufficient information about
the thermodynamic limit of form factors in massless models. The matter
is that, in distinction of massive models in infinite volume for which form factors were successfully calculated in series of
works (see e.g. \cite{KarW78,Smi90,JimML95,IzeKMT99}), 
 the thermodynamic limit of the form factors in the massless case
have a non-trivial behavior with respect to the size of the
system. This phenomenon was observed, apparently for the first time,
in \cite{Sla90} for the model of one dimensional bosons. Recently
similar results were obtained in \cite{AriKMW06} for the free
fermion limit of the XXZ spin chain. This non-trivial behavior of the
form factors with respect to the size of the system makes very
difficult their analysis directly in the infinite volume contrary to
the case of massive models.

One of the main goal of the present article is to develop a method to resolve this problem. We start with determinant representations for  form
factors for the $finite$ Heisenberg chain \cite{KitMT99}. Then, acting in the
spirit of works \cite{Sla90,IzeKMT99}, we proceed to the
thermodynamic limit of the specific form factor under consideration. This
method allows us to obtain its leading (power-law) behavior in terms of the size of the chain and to calculate the corresponding finite amplitude. Remarkably, the exponent governing the power-law decrease of the norm squared of this (umklapp-type) form factor in terms of the size $M$ of the chain is equal to the critical exponent $2{\cal Z}^2$ for the corresponding  oscillating term in  \eqref{9-corr-funct}. 

The article is organized as follows. In Section~\ref{sec-XXZ}, we
recall some well known results concerning the XXZ chain, its
algebraic Bethe ansatz solution and its thermodynamic limit. The
main result of the article is discussed in
Section~\ref{sec-results}: we define more precisely the form factor
that we investigate and we write explicitly its relation with the
amplitude appearing in \eqref{9-corr-funct}; we also recall the
exact expression of this amplitude as computed in \cite{KitKMST08b}.
The next sections are devoted to the proof of this result. In
Section~\ref{sec-ff-finite} we remind determinant
representations for this form factor in the finite chain that are suited for our analysis. It
enables us, in Section~\ref{sec-ff-lim}, to compute its thermodynamic
limit. Several technical results concerning this limit are given
in the appendices.


\section{The XXZ chain: general results}
\label{sec-XXZ}

The Hamiltonian of the integrable spin-$\tf{1}{2}$ XXZ  chain, with anisotropy parameter $\Delta$ and in an external longitudinal magnetic field $h$, is given by
\begin{equation}\label{IHamXXZ} H=H^{(0)}-hS_z, \end{equation}
where
\begin{align}\label{H0}
 &H^{(0)}=\sum_{m=1}^{M}\left\{
    \sigma^x_{m}\sigma^x_{m+1}+\sigma^y_{m}\sigma^y_{m+1}
    +\Delta(\sigma^z_{m}\sigma^z_{m+1}-1)\right\},\\
 &S_z=\frac{1}{2}\sum_{m=1}^{M}\sigma^z_{m}, \qquad
                   [H^{(0)},S_z]=0. 
   \label{Sz}
\end{align}
In this expression, as well as in \eqref{9-corr-funct}, $\sigma^{x,y,z}_{m}$ stand for the local
spin operators (in the spin-$\tf{1}{2}$ representation where they are represented by Pauli matrices) acting non-trivially
on the $m^{th}$ site of the chain.  The quantum space of states is ${\cal
H}={\otimes}_{m=1}^M {\cal H}_m$ where ${\cal H}_m\sim
\mathbb{C}^2$ is the local quantum space at site $m$.  We
assume periodic boundary conditions and, for
simplicity, the length of the chain $M$ is chosen to be even. Since the simultaneous reversal
of all spins is equivalent to a change of sign of the magnetic
field,  it is enough to consider the case $h\ge 0$.

In the thermodynamic limit ($M\to\infty$) the model exhibits different
regimes depending on the values of the anisotropy $\Delta$ and magnetic field $h$
\cite{Bax82L}. We focus on the massless regime ($|\Delta|<1$) and assume that the
magnetic field is below its critical value $h_c$ \cite{BogIK93L}. All along this article we use the
parameterization $\Delta=\cos\zeta$, $0<\zeta<\pi$.

\subsection{The finite XXZ chain in the algebraic Bethe Ansatz framework}
\label{sec-XXZ-finite}

The central object of the algebraic Bethe Ansatz method is the quantum monodromy matrix.
In the case of the XXZ chain, it is a $2\times 2$ matrix,
 \begin{equation}\label{ABAT}
 T(\lambda)=\left(\begin{array}{cc}
 A(\lambda)&B(\lambda)\\
 C(\lambda)&D(\lambda)
 \end{array}\right),
 \end{equation}
with operator-valued entries $A$, $B$, $C$ and $D$ acting on the quantum space of states
${\cal H}$ of the chain.
These operators depend on a
complex parameter $\lambda$ and satisfy a set of quadratic
commutation relations driven by the six-vertex $R$-matrix.

The Hamiltonian $H^{(0)}$ of the XXZ chain is given by a trace
identity \cite{Fad84} involving the transfer matrix
 \begin{equation}\label{Transf}
 {\cal T}(\lambda) =\tr T(\lambda)=A(\lambda)+D(\lambda).
 \end{equation}
The transfer matrices commute for arbitrary values of the spectral
parameter, $[{\cal T}(\lambda), {\cal T}(\mu)]=0$, and their common
eigenstates coincide with those of the Hamiltonian.

The solution of the quantum inverse scattering problem enables us to
express local spin operators in terms of the entries of the
monodromy matrix \cite{KitMT99,MaiT99}:
 \begin{equation}\label{FCtab}
 \sigma^\alpha_k
  ={\cal T}^{k-1} (-i\zeta/2)  
    \cdot
   \tr\big(T (-i\zeta/2)\, 
                  \sigma^\alpha\big)
    \cdot
  {\cal T}^{-k} (-i\zeta/2) . 
 \end{equation}
In the l.h.s. $\sigma^\alpha_k$ denote a local spin operator at
site $k$, whereas  $\sigma^\alpha$ appearing in the r.h.s. should
 be understood as a $2\times 2$ Pauli matrix multiplying
the $2\times 2$ monodromy matrix.

It turns out to be convenient to introduce a
slightly more general object \cite{KitMST05a}, the twisted transfer matrix
 \begin{equation}\label{Twis-T-M}
 {\cal T}_\kappa(\lambda)=A(\lambda)+\kappa D(\lambda),
 \end{equation}
depending  on an additional complex parameter $\kappa$.
For a fixed value of $\kappa$, these twisted transfer matrices also
commute with each others, $[{\cal T}_\kappa(\lambda), {\cal
T}_\kappa(\mu)]=0$,  and hence possess a common eigenbasis.
The latter goes to the one of the Hamiltonian \eqref{H0} in the  $\kappa \tend 1$ limit.

In the framework of the algebraic Bethe Ansatz, an arbitrary  quantum
state can be obtained from the states generated  by a multiple action
of $B(\lambda)$ operators  on the reference state $\ket{0}$ with all
spins up (respectively by a multiple action of $C(\lambda)$ operators
on the dual reference state $\bra{0}$). Consider the subspace
$\mathcal{H}^{(M/2-N)}$ of the space of states $\mathcal{H}$ with a
fixed number $N$ of spins down. In this subspace, the eigenstates
$\ket{\psi_\kappa(\{\mu\})}$ (respectively the dual eigenstates 
$\bra{\psi_\kappa(\{\mu\})}$) of the twisted transfer matrix ${\cal
T}_\kappa(\nu)$ can be constructed in the form
\begin{equation}\label{ABAES}
  \ket{\psi_\kappa(\{\mu\})}=\prod_{j=1}^{N}B(\mu_j)\ket{0},
  \qquad
  \bra{\psi_\kappa(\{\mu\})}=\bra{0}\prod_{j=1}^{N}C(\mu_j) \, ,
\end{equation}
where the parameters $\mu_1,\ldots,\mu_N$ satisfy the system of
twisted Bethe equations
\begin{equation}\label{TTMBE_Y}
   {\cal Y}_\kappa(\mu_j|\{\mu\})=0, \qquad j=1,\dots,N.
\end{equation}
Here, the function ${\cal Y}_\kappa$ is defined as
\begin{equation}\label{TTM_Y-def}
   {\cal Y}_\kappa(\nu|\{\mu\}) =
      a(\nu)\prod_{k=1}^{N}\sinh(\mu_k-\nu-i\zeta)
      + \kappa\,d(\nu) \prod_{k=1}^{N}\sinh(\mu_k-\nu+i\zeta),
\end{equation}
$a(\nu)$, $d(\nu)$ being the eigenvalues of the operators $A(\nu)$
and $D(\nu)$ on the reference state $\ket{0}$:
\begin{equation}\label{EV_DA}
  a(\nu) =\sinh^M\big(\nu-{\textstyle\frac{i\zeta}2}\big),\qquad
  d(\nu) =\sinh^M\big(\nu+{\textstyle\frac{i\zeta}2}\big).
\end{equation}
The corresponding eigenvalue of ${\cal T}_\kappa(\nu)$ on
$\ket{\psi_\kappa(\{\mu\})}$ (or on a dual eigenstate) is
 \begin{equation}\label{ABAEV}
 \tau_\kappa(\nu|\{\mu\})=
 a(\nu)\prod_{k=1}^{N}\frac{\sinh(\mu_k-\nu-i\zeta)}{\sinh(\mu_k-\nu)}
 + \kappa\,d(\nu)
 \prod_{k=1}^{N}\frac{\sinh(\nu-\mu_k-i\zeta)}{\sinh(\nu-\mu_k)}.
 \end{equation}
Note that all this construction is also valid at $\kappa=1$, in which case we agree upon
omitting the subscript $\kappa$ in the corresponding quantities.

Not all the solutions to the system \eqref{TTMBE_Y} yield eigenstates of the operator ${\cal
T}_\kappa(\nu)$. The detailed classification of the pertinent solutions
is given in \cite{TarV96} (see also \cite{KitMST05b}). For those
solutions corresponding to the eigenstates (the so-called admissible
solutions), the system of the twisted Bethe equations can be recast as
\begin{equation}\label{TBE-rat}
   \frac{a(\mu_j)}{d(\mu_j)}\prod_{k=1}^{N}\frac{\sinh(\mu_k-\mu_j-i\zeta)}
      {\sinh(\mu_k-\mu_j+i\zeta)}=- \kappa,
 \qquad j=1,\dots,N.
\end{equation}
%

\subsection{Thermodynamic limit}
\label{sec-XXZ-lim}

We outline here very briefly the thermodynamic limit of the model,
more detailed analysis can be found in
\cite{YanY66,LieM66,BogIK93L}.

The ground state of the infinite chain with fixed
magnetization $\langle\sigma^z\rangle$ can be constructed as the
$M\tend\infty$ limit of the finite chain ground state \eqref{ABAES} at
$\kappa=1$. Hereby the limit of the ratio $N/M$ is equal to an
average density $D$ whose value is related to the magnetization by
$\langle\sigma^z\rangle=1-2D$.

In order to construct the ground state, one first takes the
logarithm of the system \eqref{TBE-rat} at $\kappa=1$ and $N$
and $M$ fixed,
 \begin{equation}\label{BE}
 Mp_0(\mu_j)-\sum_{k=1}^N\vartheta(\mu_j-\mu_k)=2\pi n_j,
 \qquad j=1,\dots,N.
 \end{equation}
Here $-M/2<n_j\le M/2$, with $n_j\in\mathbb{Z}$ for $N$ odd and $n_j\in\mathbb{Z}+1/2$ for
$N$ even (recall that $M$ is assumed even). The functions $p_0(\lambda)$ and
$\vartheta(\lambda)$,
 \begin{equation}\label{1-p0}
 p_0(\lambda)=i\log\left(\frac{\sinh(\textstyle{\frac{i\zeta}2}+\lambda)}
 {\sinh(\textstyle{\frac{i\zeta}2}-\lambda)}\right),\qquad
 \vartheta(\lambda)=i\log\left(\frac{\sinh(i\zeta+\lambda)}
 {\sinh(i\zeta-\lambda)}\right),
 \end{equation}
are respectively called bare momentum and bare phase (we choose the principal branch of
the logarithm).
Among all the eigenstates obtained through \eqref{BE}, we define the $N$-particle
ground state to be the state with the lowest energy in the sector with
$N$ spins down. It is characterized by the (half-)integers
$n_j=I_j\equiv j-(N+1)/2$ \cite{YanY66}, and the corresponding roots
$\lambda_j$ are real numbers belonging to some interval $[-q,q]$
called the Fermi zone.

At the next step the length of the chain $M$ as well as the number
of spins down $N$ are sent to infinity at fixed limiting value of
$N/M$.  In this limit the Bethe roots $\lambda_j$ corresponding to the
$N$-particle ground state condensate,
$\lambda_{j+1}-\lambda_j=O(M^{-1})$ (see e.g. \cite{LieM66}), and the
system \eqref{BE} turns into a linear integral equation for the
spectral density $\rho(\lambda)$ \footnote{Although there are evidences that the ground state can be described in that way, strictly speaking, to our knowledge, this statement has been proved only in the case  $\Delta \le 0$ \cite{YanY66,LieM66}.}.

A standard way to derive this
integral equation is to introduce a ground state counting function
$\hat\xi(\lambda)$. Namely, let $\lambda(\hat\xi)$ satisfy the functional equation
 \begin{equation}\label{BE-xi}
 p_0\bigl(\lambda(\hat\xi)\bigr)-\frac1M\sum_{k=1}^N\vartheta\bigl(\lambda(\hat\xi)-\lambda_k\bigr)=2\pi
 \left(\hat\xi-\frac{N+1}{2M}\right),
 \end{equation}
for a given value of $\hat\xi$ and with parameters $\lambda_k$ corresponding to the $N$-particle ground
state. Evidently, $\hat\xi(\lambda)$ is the counting function, since
$\lambda(\frac jM)=\lambda_j$ and $\hat\xi(\lambda_j)=j/M$ due to
equation \eqref{BE}. The (finite-size) density of the $N$-particle
ground state is defined by
$\hat\rho(\lambda)=\tf{\dd\hat\xi}{\dd\lambda}$. From \eqref{BE-xi},
it is equal to
 \begin{equation}\label{Be-xi1}
 2\pi \hat\rho(\lambda)
 =p'_0(\lambda)-\frac1M\sum_{k=1}^N K(\lambda-\lambda_k),
 \end{equation}
with
 \begin{equation}\label{0-K-XXZ}
 K(\lambda)=\vartheta'(\la)=\frac{\sin2\zeta}{\sinh(\lambda+i\zeta)\sinh(\lambda-i\zeta)}.
 \end{equation}

In the thermodynamic limit, the $N$-particle density
$\hat\rho(\lambda)$  goes to the spectral density $\rho(\lambda)$ of
the infinite chain with support on a finite interval $[-q,q]$. In
this limit, discrete sums of piecewise continuous functions can be
replaced by integrals \textit{via} the Euler--Maclaurin summation formula
 \begin{equation}\label{sum-int}
 \lim_{N,M\to\infty}\frac1M\sum_{k=1}^N f(\lambda_k)
 =\lim_{N,M\to\infty}\frac1M\sum_{k=1}^N
 f\bigl(\hat\xi^{-1}({\textstyle\frac
 kM})\bigr)= \int\limits_{-q}^q f(\lambda) \, \rho(\lambda)\, \dd\lambda.
 \end{equation}
Thus, replacing the sum over $k$ by an integral in \eqref{Be-xi1}, we
obtain the integral equation for the density of the infinite chain,
 \begin{equation}\label{0-rho}
 \rho(\lambda)+\frac1{2\pi}\int\limits_{-q}^q K(\lambda-\mu) \, \rho(\mu)\, \dd\mu
 =\frac1{2\pi}p'_0(\lambda).
 \end{equation}
The average density $D=\lim_{N,M\to\infty}N/M$ is then given by
 \begin{equation}\label{D-int}
 D=\lim_{N,M\to\infty}\frac1M\sum_{k=1}^N 1=
 \int\limits_{-q}^q  \rho(\lambda)\, \dd\lambda,
 \end{equation}
which defines the integration boundary $q$ in \eqref{0-rho}.

Observe that the definition of the discrete density
$\hat\rho(\lambda)$ implies that $\rho(\la)$ can also be defined as
the limiting value
 \begin{equation}\label{rho-lim}
 \rho(\lambda_j)=\lim_{N,M\to\infty}\frac{\hat\xi(\lambda_{j+1})-\hat\xi(\lambda_j)}{\lambda_{j+1}-\lambda_j}
 =\lim_{N,M\to\infty}\frac1{M(\lambda_{j+1}-\lambda_j)}.
 \end{equation}

The method described above allows one to construct the ground state
of the infinite XXZ chain at fixed magnetization. For the XXZ chain
in an external magnetic field, the magnetization of the ground state
is not an independent variable but depends on the magnetic field $h$.
Then the boundary of the Fermi zone $q$ in the equation
\eqref{0-rho} is defined from the condition $\varepsilon(q)=0$ (instead of from \eqref{D-int}),
where $\veps(\lambda)$ is the dressed energy satisfying an integral
equation similar to \eqref{0-rho}:
 \begin{equation}\label{energy}
 \varepsilon(\lambda)
 +\frac{1}{2\pi}\int\limits_{-q}^{q}K(\lambda-\mu) \, \varepsilon(\mu) \, \dd\mu
 =h-2p'_0(\lambda)\sin \zeta\; .
 \end{equation}
The condition $\varepsilon(q)=0$ provides the positiveness of the
energy of any excited state. We see from \eqref{energy} that the Fermi boundary $q$ depends on
the anisotropy parameter $\Delta$ and on the magnetic field $h$,
{\it i.e.} that $q$ is a function of the parameters of the Hamiltonian
\eqref{IHamXXZ}. We remind that $q$ remains finite for non-zero
magnetic fields, while it tends to infinity when $h\tend 0$.
Throughout the present article, $q$ will be kept finite in all the
computations.

Let us finally recall the definitions of the dressed momentum and charge, which are two important functions characterizing the ground state.
The dressed momentum is closely related to the density as
 \begin{equation}\label{0-Dmom}
 p(\lambda)=2\pi\int\limits_0^\lambda \rho(\mu)\,\dd\mu.
 \end{equation}
Obviously, the thermodynamic limit $\xi(\lambda)$ of the counting
function $\hat\xi(\lambda)$ can be expressed in terms of the dressed
momentum according to $\xi(\lambda)=[p(\lambda)+p(q)]/2\pi$.
The dressed charge satisfies the integral equation
 \begin{equation}\label{0-DC}
 Z(\lambda)+\frac1{2\pi}\int\limits_{-q}^qK(\lambda-\mu)\; Z(\mu)\,\dd\mu=1,
 \end{equation}
and can be interpreted in the XXZ model as the intrinsic magnetic
moment of the elementary excitations \cite{BogIK93L}. The quantities
${\cal Z}=Z(q)$ and $p_{{}_F}=p(q)=\pi D$ (Fermi momentum) enter the
asymptotic formula \eqref{9-corr-funct}.

The excitations above the ground state can be constructed by
adjoining particles and holes to the Dirac sea
\cite{Bet31,Hul38,Orb58,Wal59,CloG66,YanY66,BabdVV83}.
 In other words, an excited state above the ground state
corresponds to the replacement of a finite number of the
(half-)integers $I_j=j-(N+1)/2$ parameterizing the ground state by
some different ones, not belonging to the original sequence. In
general, this induces a shift of the Bethe parameters $\{\mu\}$
describing this excited state with respect to the ground state Bethe
roots $\{\la\}$, and  possible appearance of complex roots.

\section{Statement of the result}
\label{sec-results}

We are now in position to be more precise about the result we prove in this article.

We  study a particular form factor of the $\sg^z$ operator,
involving  a very specific excited state, namely the one with one
particle and one hole located on the opposite ends of the Fermi
zone. For definiteness, we consider the case where the spectral
parameters of the particle and hole are  respectively equal to
$\pm q$. This means that the (half-)integers which characterize this
state through \eqref{BE}, that we will denote $I'_j$
($j=1,\ldots,N$), coincide with the ground state ones ($I'_j=I_j$)
for $j=2,\dots,N$, but differ at $j=1$: $I'_1=N+1-(N+1)/2$. There
exists another way of describing the same eigenstate: by shifting the
ground state (half-)integers by $1$, \textit{i.e.} by setting
$I'_j=j+1-(N+1)/2$. In the framework of this approach, we can
consider the aforementioned excited state as a limit of a {\it
twisted ground state}. Indeed, similarly to \eqref{BE} different
eigenstates of the twisted transfer matrix are characterized by sets
of (half-)integers $\tilde{n}_j$ in the logarithmic form of the twisted Bethe
equations \eqref{TBE-rat},
 \begin{equation}\label{TBE}
 Mp_0(\mu_j)-\sum_{k=1}^N\vartheta(\mu_j-\mu_k)=2\pi \tilde{n}_j-i\beta,
 \qquad j=1,\dots,N,
 \end{equation}
where $\kappa=e^\beta$. In the case $\tilde{n}_j=I_j$, we call the
corresponding eigenstate the twisted ground state. In that case and at
$\beta=2\pi i$, the equations \eqref{TBE} reduce to \eqref{BE} with
$n_j=I'_j=j+1-(N+1)/2$, and, thus, the twisted ground state becomes
the excited state described above.

\bigskip

Our result is the following :

\bigskip

{\em In the thermodynamic limit, the normalized form factor of the $\sg^z$
operator between the ground state $\ket{\psi(\{\la\})}$ and the excited state $\bra{\psi(\{\mu\})}$ defined above is related to the amplitude  $|F_\sigma|^2$ (which is a finite number)
appearing in the asymptotic formula \eqref{9-corr-funct} for the longitudinal spin-spin correlation function
as
 \begin{equation}\label{m-el-sigmaA}
 \lim_{N,M\to\infty}\left(\frac{M}{2\pi}\right)^{2{\cal Z}^2}\frac{|\langle\psi(\{\mu\})|\sigma^z| \psi(\{\lambda\})\rangle|^2}
 {\|\psi(\{\mu\})\|^2\cdot\|\psi(\{\lambda\})\|^2}
 =|F_\sigma|^2.
 \end{equation}
In other words, it means that the modulus squared of the corresponding form factor behaves as $ |F_\sigma|^2 \left(\frac{M}{2\pi}\right)^{-2{\cal Z}^2}\  $ for $M$ large. It is remarkable that the exponent $2{\cal Z}^2$ governing the modulus squared of the form factor power-law decrease in terms of the size $M$ of the chain is exactly equal to the exponent for the power-law behavior of the corresponding oscillating term in the correlation function \eqref{9-corr-funct}, but there in terms of the distance $m$ separating the two spin operators.  
}

\bigskip

This result will be proven throughout Sections~\ref{sec-ff-finite} and
\ref{sec-ff-lim}. There, the above special form factor will be
computed and its thermodynamic limit will be taken.

In order to complete this statement, let us recall the explicit value of the amplitude $|F_\sigma|^2$ obtained in \cite{KitKMST08b}:
 \begin{equation}\label{7-corr-funct-1}
 |F_\sigma|^2
 = \left(\frac{2G(2,{\cal Z}) \, \sin p_{{}_F}}{\pi{\cal Z}}\right)^2
     \cdot
     \bigl[ 2\pi \, \rho(q) \, \sinh(2q)\bigr]^{-2{\cal Z}^2}
     \cdot
     e^{C_1-C_0}
     \cdot
     \bigl.\widetilde{\cal A}(\beta)\bigr|_{\beta=2\pi i}.
 \end{equation}
Here the notation $G(a,x)$ represents the product of Barnes
functions $G(a,x)=G(a+x)G(a-x)$ (\textit{cf} \ref{Appendix Barnes
Function}). The constants $C_0$ and $C_1$ are defined in terms of
the dressed charge $Z$:
 \begin{equation}\label{7-C0}
 C_0={\dis\int\limits_{-q}^{q}}\frac{Z(\lambda) \, Z(\mu)}{\sinh^2(\lambda-\mu-i\zeta)}\,
      \dd\lambda \, \dd\mu\, ,
 \end{equation}
 \vspace{-4mm}
 \begin{equation}\label{7-C_1}
 C_1=\frac12\int\limits_{-q}^{q}
           \frac{Z'(\lambda)\, Z(\mu) -Z(\lambda)\, Z'(\mu)}{\tanh(\lambda-\mu)}\,
           \dd\lambda \, \dd\mu
           +2{\cal Z}\int\limits_{-q}^{q}\frac{{\cal Z}-Z(\lambda)}{\tanh(q-\lambda)}\, \dd\lambda \, .
 \end{equation}
Finally, the coefficient $\widetilde{\cal A}(\beta)$ is given in terms of a ratio of Fredholm determinants:
 \begin{equation}\label{7-const-A}
 \widetilde{\cal A}(\beta)=
 \left| \,
  e^{\frac\beta2 [ \tilde z(q-i\zeta)-\tilde z(-q-i\zeta) ]} \;
 \frac{\det\big[ I+\frac1{2\pi i}U_{-q}^{(\lambda)}(w,w') \big]}
         {\det\big[ I+\frac1{2\pi }K \big]} \,
 \right|^2.
 \end{equation}
In this expression, the integral
operator $I+\tf{K}{2\pi}$ occurring in the denominator acts on the interval $[-q,q]$ with kernel
\eqref{0-K-XXZ}, whereas the integral operator appearing in the numerator acts on a contour surrounding  $[-q,q]$ with kernel
 \begin{equation}\label{7-U1}
 U_{-q}^{(\lambda)}(w,w')
 =-\frac{ e^{\beta\tilde z(w)} \,   [ K_\kappa(w-w')-K_\kappa(-q-w') ]  }
            { e^{\beta\tilde z(w+i\zeta)} - e^{\beta+\beta\tilde z(w-i\zeta)} },
 \end{equation}
where
 \begin{equation}\label{2-Kk}
  K_\kappa(\lambda)=\coth(\lambda+i\zeta)-\kappa\coth(\lambda-i\zeta),
  \qquad  \kappa=e^\beta.
 \end{equation}
The function $\tilde z(w)$ appearing \eqref{7-const-A}, \eqref{7-U1}, is the $i\pi$-periodic Cauchy
transform of the dressed charge $Z$,
\begin{equation}\label{9-CauchyZ}
 \tilde{z}(w)
 = \frac{1}{2\pi i} \int\limits_{-q}^{q} \coth(\lambda-w) \, Z(\lambda) \, \dd\lambda.
 \end{equation}
Observe that it has a cut on the interval $[-q,q]$ and satisfies the following properties:
 \begin{align}
 &\tilde z(w+i\zeta)- \tilde z(w-i\zeta)=1-Z(w) \, ,
   \label{6-prop-z1}   \\
 &\tilde z_+(w)- \tilde z_-(w)=Z(w)\, , \qquad w\in[-q,q] \, .
  \label{6-prop-z2}
\end{align}
The equation \eqref{6-prop-z1} follows from \eqref{0-DC}. In
\eqref{6-prop-z2}, $\tilde z_\pm$ stand for the limiting values of
$\tilde z$ when $\R$ is approached from the upper and lower half
planes.

\begin{rem}
Equation \eqref{7-corr-funct-1} contains the limiting value of
$\widetilde{\cal A}(\beta)$ at $\beta=2\pi i$. We did not set
$\beta=2\pi i$ directly in \eqref{7-const-A}--\eqref{2-Kk}, as it
may cause problems at some particular values of $\De$. For example,
at $\Delta=0$ ($\zeta=\frac\pi2$, $Z(\lambda)=1$), the kernel
\eqref{7-U1} becomes ill-defined when $\beta=2\pi i$. It should therefore be understood in the sense of the limit $\be \tend
2i\pi$, the latter being well defined.
\end{rem}


\section{Special form factor for the finite chain}
\label{sec-ff-finite}

The solution of the quantum inverse scattering problem \eqref{FCtab}
and determinant representations for scalar products
\cite{Sla89,KitMT99} give a possibility to express, for the finite chain,
arbitrary form factors of local spin operators as finite-size
determinants \cite{KitMT99}. We explain here how to obtain such a
representation for the special form factor defined above. Then,
following the ideas of \cite{IzeKMT99,KitKMST08b}, we transform the
obtained determinant to a new one, more convenient for the calculation of
the thermodynamic limit.

From now on, we will consider eigenstates of the twisted transfer
matrix only for $\kappa$ belonging to the unit circle, {\it i.e.}
$\kappa=e^\beta$, where $\beta$ is pure imaginary. This restriction
is not crucial, but convenient. The matter is that, due to the
involution $A^\dagger(\bar\lambda)=D(\lambda)$  ($\dagger$ means
Hermitian conjugation), the operator $e^{-\beta/2}{\cal
T}_\kappa(\lambda)$ becomes self-adjoint for $|\kappa|=1$ and
$\lambda\in\mathbb{R}$. In such a case, the roots of the twisted Bethe
equations \eqref{TBE} are real or contain complex conjugated pairs. In its turn,
 the involution $B^\dagger(\bar\lambda)=-C(\lambda)$ guarantees the duality of eigenstates $\langle \psi_\kappa(\{\mu\})|=(-1)^N
|\psi_\kappa(\{\mu\})\rangle^\dagger$.

\subsection{Special form factor and scalar product}

Consider the following operator ${\cal Q}_m$, giving the number of spins down in
the first $m$ sites of the chain:
 \begin{equation}\label{0-Q1}
 {\cal Q}_m=\frac12\sum_{n=1}^{m}(1-\sigma_n^z)\,  .
 \end{equation}
It is easy to see that
 \begin{equation}\label{0-defQ}
 e^{\beta {\cal Q}_m}=\prod_{n=1}^m\left(\frac{1+\kappa}2+\frac{1-\kappa}2\cdot
 \sigma_n^z\right),
 \qquad \kappa=e^\beta,
 \end{equation}
and we have
 \begin{equation}\label{0-corr-funct}
 2D_m\left.\frac{\partial}{\partial\beta}
 e^{\beta {\cal Q}_m}\right|_{\beta=2\pi in}=1-\sigma_{m+1}^z,\qquad
 n\in\mathbb{Z},
 \end{equation}
where the symbol $D_m$ stands for the  lattice derivative:
$D_mf(m)=f(m+1)-f(m)$.
Also, note  that $e^{\beta {\cal Q}_m}$ is a polynomial in $\kappa$ which becomes the identity operator whenever $\beta=2\pi in$,
$n\in\mathbb{Z}$.

The operator \eqref{0-defQ} admits, through the solution of the inverse problem \eqref{FCtab},
a simple representation in terms
of the twisted and standard transfer matrices:
 \begin{equation}\label{Q-T}
 e^{\beta {\cal Q}_m}
 ={\cal T}_\kappa^{m}\big(-{\textstyle\frac{i\zeta}2}\big) \;
 {\cal T}^{-m}\big(-{\textstyle\frac{i\zeta}2}\big).
 \end{equation}
We use this representation and consider the following
matrix element of $e^{\beta {\cal Q}_m}$:
 \begin{equation}\label{m-el}
 \langle\psi_\kappa(\{\mu\})|e^{\beta {\cal Q}_m}|
 \psi(\{\lambda\})\rangle
 =\langle\psi_\kappa(\{\mu\})| {\cal T}_\kappa^{m}\big(-{\textstyle\frac{i\zeta}2}\big)\;
 {\cal T}^{-m}\big(-{\textstyle\frac{i\zeta}2}\big)|
 \psi(\{\lambda\})\rangle,
 \end{equation}
where $| \psi(\{\lambda\})\rangle$ is the ground state in the
$N$-particle sector, and $\langle\psi_\kappa(\{\mu\})|$ is an
eigenstate of the twisted transfer matrix ${\cal T}_\kappa$.  Then
 \begin{equation}\label{m-el1}
 \langle\psi_\kappa(\{\mu\})|e^{\beta {\cal Q}_m}|
 \psi(\{\lambda\})\rangle=
 \bigg(\frac{\tau_\kappa(\frac{-i\zeta}2|\{\mu\})}{\tau(\frac{-i\zeta}2|\{\lambda\})}\bigg)^m
 \langle\psi_\kappa(\{\mu\})| \psi(\{\lambda\})\rangle.
 \end{equation}
Setting $\nu=-\frac{i\zeta}2$ in \eqref{ABAEV}, we obtain
 \begin{equation}\label{m-el2}
 \langle\psi_\kappa(\{\mu\})|e^{\beta {\cal Q}_m}|
 \psi(\{\lambda\})\rangle
 =e^{im\sum_{j=1}^{N} [p_0(\mu_j)-p_0(\lambda_j) ]}
 \langle\psi_\kappa(\{\mu\})| \psi(\{\lambda\})\rangle.
 \end{equation}

Let us now differentiate \eqref{m-el2} with respect to $\beta$ at
$\beta=2\pi i$. Hereby, at $\beta=2\pi i$, the state
$\langle\psi_\kappa(\{\mu\})|$ becomes an eigenstate of the standard
transfer matrix.
As it does not coincide with the
ground state, it is orthogonal to the last one, and thus,
 \begin{equation}\label{m-el-Q1}
 \langle\psi(\{\mu\})|{\cal Q}_m| \psi(\{\lambda\})\rangle
 =\left(
 e^{im\sum_{j=1}^{N}[p_0(\mu_j)-p_0(\lambda_j)]}-1\right)
 \Bigl.\frac{\partial}{\partial\beta}\langle\psi_\kappa(\{\mu\})|
 \psi(\{\lambda\})\rangle\Bigr|_{\beta=2\pi i}.
 \end{equation}
Using also \eqref{0-corr-funct}  we find that the special form
factor we want to compute can be obtained in terms of the
normalized scalar product between the ground state
$\ket{\psi(\{\la\})}$ and the twisted ground state
$\bra{\psi_\kappa(\{\mu\})}$ given by \eqref{TBE} with $\tilde{n}_j=I_j$:
 \begin{equation}\label{m-el-sigma}
 \frac{|\langle\psi(\{\mu\})|\sigma^z_{m+1}| \psi(\{\lambda\})\rangle|^2}
 {\|\psi(\{\mu\})\|^2\cdot\|\psi(\{\lambda\})\|^2}
 =-8\sin^2\left(\frac{{\cal P}_{ex}}2\right) \cdot
  \left.\frac{\partial^2}{\partial\beta^2}
 S_N^{(\kappa)}(\{\mu\},\{\lambda\})
 \right|_{\beta=2\pi i},
 \end{equation}
with
 \begin{equation}\label{S}
 S_N^{(\kappa)}(\{\mu\},\{\lambda\})
 =\left(\frac{ \langle\psi_\kappa(\{\mu\})|
 \psi(\{\lambda\})\rangle   }
 {\|\psi_{\kappa}(\{\mu\})\|\cdot\|\psi(\{\lambda\})\|}\right)^2,
 \end{equation}
and where ${\cal P}_{ex}=\sum_{j=1}^N [p_0(\mu_j)-p_0(\lambda_j) ]$
is the total momentum of the excitation. Deriving \eqref{m-el-sigma}
we have used that
$\langle\psi_\kappa(\{\mu\})|\psi(\{\lambda\})\rangle$ is real at
$|\kappa|=1$.

In \eqref{S}, the parameters $\mu_j$ satisfy the system of twisted
Bethe equations \eqref{TBE} with $\tilde{n}_j=I_j$ and, hence, are functions
of $\beta$. In particular, they are taken to be
 \begin{equation}\label{choice-mu}
 \mu_j(\beta)\bigl.\bigr|_{\beta=0}=\lambda_j, \qquad j=1,\dots,N.
 \end{equation}
Then, in the l.h.s. of \eqref{m-el-sigma}, the set
$\paa{\mu_j(\beta=2\pi i)}$ describes the excited state considered
in Section~\ref{sec-results} with a particle at $q$ and a hole at
$-q$.

\subsection{Determinant representation}
\label{sSP}

Thus, in order to study our special form factor, we can use the determinant representations
for the scalar product between an
eigenstate of the twisted transfer matrix with any arbitrary state
of the form \eqref{ABAES}.
Let us recall these representations.

\begin{prop}\cite{Sla89,KitMT99}
Let $\mu_1,\dots,\mu_N$ satisfy the system \eqref{TTMBE_Y} and
$\lambda_1,\dots,\lambda_N$ be generic complex numbers.
Then
\begin{align}
\langle 0|\prod_{j=1}^{N}C(\lambda_j)|\psi_\kappa(\{\mu\})\rangle &=
\langle\psi_\kappa(\{\mu\})|\prod_{j=1}^{N}B(\lambda_j)|0\rangle
\hspace{5.5cm}\nonumber
\\
&=  \frac{\prod_{a=1}^{N} d(\mu_a)}
{\prod\limits_{a>b}^N\sinh(\mu_a-\mu_b)\sinh(\lambda_b-\lambda_a)}
\cdot \det_N
\Omega_\kappa(\{\mu\},\{\lambda\}|\{\mu\})\label{scal-prod},
\end{align}
where  the $N\times N$ matrix
$\Omega_\kappa(\{\mu\},\{\lambda\}|\{\mu\})$  is defined as
\begin{multline} \label{matH}
  (\Omega_\kappa)_{jk}(\{\mu\},\{\lambda\}|\{\mu\})=
  a(\lambda_k)\,t(\mu_j,\lambda_k)\,\prod_{a=1}^{N} \sinh(\mu_a-\lambda_k-i\zeta)\\
   -\kappa\, d(\lambda_k)\,t(\lambda_k,\mu_j)\,\prod_{a=1}^{N} \sinh(\mu_a-\lambda_k+i\zeta),
\end{multline}
with
\begin{equation}\label{def-t}
t(\mu,\lambda)=\frac{-i\sin\zeta}{\sinh(\mu-\lambda)\sinh(\mu-\lambda-i\zeta)}.
\end{equation}
\end{prop}

Note that, due to the Bethe equations \eqref{TTMBE_Y}, the entries
of the matrix $\Omega_\kappa\pa{\paa{\mu},\paa{\la} \mid \paa{\mu}
}$ are not singular at $\lambda_k=\mu_j$. In particular, if
$\lambda_j=\mu_j$ for all $j=1,\dots,N$, then we obtain the square
of the norm of the twisted eigenstate $\langle
\psi_\kappa(\{\mu\})|\psi_\kappa(\{\mu\})\rangle$ (recall that
$|\kappa|=1$), which reduces to
 \begin{multline}\label{norm-Theta}
 \langle \psi_\kappa(\{\mu\})|\psi_\kappa(\{\mu\})\rangle
 = (-1)^N
 \prod_{j=1}^{N}  \bigl[2\pi iM \, \hat\rho_\kappa(\mu_j) \, a(\mu_j) \, d(\mu_j)\bigr]
 \frac{\prod\limits_{a,b=1}^N\sinh(\mu_{a}-\mu_b-i\zeta)}
        {\prod\limits_{a,b=1\atop{a\ne b}}^N\sinh(\mu_a-\mu_b)}
   \\
   \times
 \det_N \Theta_{jk}^{(\mu)},
 \end{multline}
with
 \begin{equation}\label{2-Gjk}
 \Theta_{jk}^{(\mu)}=\delta_{jk}+\frac{K(\mu_j-\mu_k)}{2\pi M\hat\rho_\kappa(\mu_k)},
 \qquad
 2\pi M \hat\rho_\kappa(\mu)=-i\log'\frac{a(\mu)}{d(\mu)}-\sum_{a=1}^N K(\mu-\mu_a).
 \end{equation}

In the case of interest (scalar product of the untwisted ground
state $\ket{ \psi(\{\lambda\}) }$ with the $\kappa$-twisted one
$\bra{ \psi_\kappa(\{\mu\}) }$), one has two different
representations for the scalar product. On the one hand, it is
possible to apply equation \eqref{scal-prod} when $\ket{
\psi(\{\lambda\}) }$ in understood as an arbitrary state. On the
other hand, one can consider $\ket{ \psi(\{\lambda\}) }$ as an
eigenstate (for $\kappa=1$), interpret $\bra{\psi_\kappa(\{\mu\}) }$
as an arbitrary state, and apply formula \eqref{scal-prod} with
$\lambda\leftrightarrow\mu$. Thus, one can obtain two different, but
equivalent representations for the scalar product
$\bra{\psi_\kappa(\{\mu\}) }\psi(\{\lambda\})\rangle$. We do not
give here the explicit formulae (which can be easily derived from \eqref{scal-prod}--\eqref{def-t}) 
as these representations
are not convenient for the calculation of  thermodynamic
limits. In order to obtain new formulae that are appropriate for such
goal, one can extract the products of $\sinh^{-1}(\mu_k-\lambda_j)$
from the determinant of the matrix $\Omega_\kappa$ (see
\cite{KitKMST08b,Sla90,IzeKMT99}). As a result, one obtains two new
representations containing  Fredholm determinants of integral
operators of the form $I+\frac1{2\pi i}\widehat
U_\theta^{(\lambda,\mu)}(w,w')$, with kernels
 \begin{align}\label{2-U1}
& \widehat U_\theta^{(\lambda)}(w,w')=
 -\prod\limits_{a=1}^N\frac{\sinh(w-\mu_a)}{\sinh(w-\lambda_a)}\cdot
 \frac{K_\kappa(w-w')-K_\kappa(\theta-w')}{V_+^{-1}(w)-\kappa
 V_-^{-1}(w)},
   \\
   \label{2-U2}
 & \widehat U_\theta^{(\mu)}(w,w')=
 \prod\limits_{a=1}^N\frac{\sinh(w'-\lambda_a)}{\sinh(w'-\mu_a)}\cdot
 \frac{K_\kappa(w-w')-K_\kappa(w-\theta)}{V_-(w')-\kappa
 V_+(w')}.
 \end{align}
Here  $\theta$ is an arbitrary complex number, $K_\kappa$ is given by \eqref{2-Kk}, and
 \begin{equation}\label{1-Vpm}
 V_{\pm}(w)\equiv
 V_{\pm} \left( w \mid \begin{matrix} \{\lambda\}_N \\ \{\mu\}_N \end{matrix}\right)
 = \prod_{a=1}^{N}
 \frac {\sinh(w-\lambda_a\pm i\zeta)}{\sinh(w-\mu_a\pm i\zeta)}.
 \end{equation}
These integral operators act on a contour surrounding the points
$\{\lambda\}$ (resp.$\{\mu\}$) but no any other singularity of
$\widehat U_{\th}^{(\lambda,\mu)} \pa{w,w'}$. More precisely, one
has the following proposition (see Appendix A of \cite{KitKMST08b}
for the proof):

\begin{prop}\label{Extract-Cauchy} \cite{KitKMST08b}
Let $\lambda_1,\dots,\lambda_N$ satisfy the system of
untwisted Bethe equations and $\mu_1,\dots,\mu_N$ satisfy the
system of $\kappa$-twisted Bethe equations \eqref{TTMBE_Y}.
Then
 \begin{multline}\label{2-l-rep}
 \langle \psi(\{\lambda\})|\psi_\kappa(\{\mu\})\rangle
 =\prod_{a,b=1}^N
 \frac{\sinh(\mu_a-\lambda_b-i\zeta)}{\sinh(\lambda_a-\mu_b)}\cdot
 \prod_{j=1}^N\left\{d(\mu_j)\, d(\lambda_j)
      \left[\kappa \frac{V_+(\lambda_j)}{V_-(\lambda_j)}-1\right]\right\}
  \num
 \times
  \frac{1-\kappa}{V_+^{-1}(\theta)-\kappa V_-^{-1}(\theta)}\cdot
  \det\left[ I+\frac1{2\pi i} \widehat U_\theta^{(\lambda)}(w,w') \right] ,
  \end{multline}
and
 \begin{multline}\label{2-z-rep}
%
 \langle \psi(\{\lambda\})|\psi_\kappa(\{\mu\})\rangle=
 \prod_{a,b=1}^N\frac{\sinh(\lambda_a-\mu_b-i\zeta)}{\sinh(\mu_a-\lambda_b)}\cdot
 \prod_{j=1}^N\left\{d(\mu_j)\, d(\lambda_j)\left[\frac{V_-(\mu_j)}{V_+(\mu_j)}-\kappa\right]\right\}
  \num
 \times
  \frac{1-\kappa}{V_-(\theta)-\kappa V_+(\theta)}\cdot
    \det\left[ I+\frac1{2\pi i}\widehat U_\theta^{(\mu)}(w,w')\right] .
  \end{multline}
\end{prop}

We would like to emphasize that equations \eqref{2-l-rep}  and
\eqref{2-z-rep} describe scalar products for the {\em finite} XXZ
chain even though they contain Fredholm determinants of integral
operators. In fact, it is not difficult to see that the integral
operators $\widehat U_{\th}^{(\lambda,\mu)}$ are of finite rank and
hence the Fredholm determinants can be reduced to usual determinants
of $N\times N$ matrices \cite{KitKMST08b}.

\begin{rem}
We gave here two different representations for the same quantity,
because we will use both of them. Namely, we will substitute
\eqref{2-l-rep} for one scalar product $\langle\psi_\kappa(\{\mu\})|
\psi(\{\lambda\})\rangle$ in \eqref{S} and  \eqref{2-z-rep} for the
other one, despite the evident fact that these scalar products are the same. This way
seems to be slightly strange, but it leads directly to equation
\eqref{7-corr-funct-1}. Due to the equivalence of the
representations \eqref{2-l-rep} and \eqref{2-z-rep} one can use, of
course, only one of them, coming eventually to the same result. The
last way, however, requires some additional identities, since
the equivalence of the representations \eqref{2-l-rep} and
\eqref{2-z-rep} is non-trivial.
\end{rem}

It remains to apply all these results to our special form factor,
which was expressed in terms of the normalized scalar product $S_N^{(\kappa)}$ through \eqref{m-el-sigma}.
Substituting \eqref{2-l-rep}, \eqref{2-z-rep}, \eqref{norm-Theta} in \eqref{S}, and
using the Bethe equations \eqref{TBE-rat}, we obtain a representation for
$S_N^{(\kappa)}$ as a product of three factors:
 \begin{equation}\label{S-prom}
 S_N^{(\kappa)}(\{\mu\},\{\lambda\})=
 {\cal A}_N^{(\kappa)}\cdot D_N^{(\kappa)}\cdot \exp C_{0,N}^{(\kappa)}  \, ,
 \end{equation}
with
 \begin{equation}\label{6-DNk}
 D_N^{(\kappa)}=\left(\det_N\frac1{\sinh(\mu_j-\lambda_k)}\right)^2\cdot
 \prod\limits_{j=1}^N\frac{\left(\frac{\kappa
 V_+(\lambda_j)}{V_-(\lambda_j)}-1\right)\left(\frac{V_-(\mu_j)}{\kappa
 V_+(\mu_j)}-1\right)}
 {4\pi^2M^2 \, \hat\rho(\lambda_j) \, \hat\rho_\kappa(\mu_j)},
 \end{equation}
 \begin{equation}\label{W}
 C_{0,N}^{(\kappa)}
 =\sum_{a,b=1}^N\log
   \frac{\sinh(\lambda_a-\mu_b-i\zeta) \, \sinh(\mu_b-\lambda_a-i\zeta)}
           {\sinh(\lambda_a-\lambda_b-i\zeta) \, \sinh(\mu_a-\mu_b-i\zeta)},
 \end{equation}
and
 \begin{equation}\label{6-ev-phiC}
 {\cal A}_N^{(\kappa)}
    =\frac{(\kappa-1)^2 \,
                \det\Big[I+\frac1{2\pi i}\widehat U^{(\lambda)}_{\theta_1}(w,w')\Big] \,
                \det\Big[I+\frac1{2\pi i}\widehat U^{(\mu)}_{\theta_2}(w,w')\Big]
            }{\left(V_+^{-1}(\theta_1)-\kappa  V_-^{-1}(\theta_1)\right)
              \bigl(V_-(\theta_2)-\kappa  V_+(\theta_2)\bigr)
              \cdot   \det_N\Theta_{jk}^{(\lambda)}
              \cdot   \det_N\Theta_{jk}^{(\mu)}}.
 \end{equation}
Recall that the representations \eqref{2-l-rep}, \eqref{2-z-rep} contain an
arbitrary complex parameter $\theta$.  Here we have used two
different parameters $\theta_1$ and $\theta_2$ for each of the scalar
products. In the following, we shall set $\theta_1=-q$ and $\theta_2=q$.

\section{Thermodynamic limit of the special form factor}
\label{sec-ff-lim}

We  compute now the large size behavior ($N,M\to\infty$, $N/M\tend D$) of the special
form factor represented in \eqref{m-el-sigma}, \eqref{S-prom}. We
need first to characterize how the parameters $\mu_j$ of the
$\kappa$-twisted ground state Bethe Ansatz equations behave in this limit. More
precisely, how they are shifted with respect to the parameters
$\lambda_j$ of the untwisted ground state. This will enable us to
take the thermodynamic limit in the expression \eqref{S-prom} for
the normalized scalar product. Whereas it is quite straightforward
to evaluate the limit of the factors ${\cal A}_N^{(\kappa)}$ and
$C_{0,N}^{(\kappa)}$ (see Subsection~\ref{sec-A-C}), we will see in
Subsection~\ref{sec-D} that the factor $D_N^{(\kappa)}$
\eqref{6-DNk} requires more attention.

\subsection{Twisted ground state and shift functions}
\label{sec-shift}

In order to study the $\kappa$-twisted ground state $\ket{\psi_\kappa(\{\mu\})}$ in the thermodynamic limit, it is convenient to introduce,
similarly as for the untwisted case (see \eqref{BE-xi}), a twisted counting function $\hat\xi_\kappa(\mu)$
satisfying $\hat\xi_\kappa(\mu_j)=j/M$. Its explicit expression is given by
 \begin{equation}\label{TBE-cf}
 \hat\xi_\kappa(\omega)
    =\frac1{2\pi}p_0(\omega)-\frac1{2\pi M}\sum_{k=1}^N\vartheta(\omega-\mu_k)
    +\frac{N+1}{2M}+\frac{i\beta}{2\pi M} \, ,
%
  \end{equation}
with $p_0$ and $\vartheta$ given by \eqref{1-p0}.
The corresponding discrete $N$-particle density
$\hat\rho_\kappa(\mu)=\frac{\dd\hat\xi_\kappa}{\dd\mu}$ coincides
with the function defined in \eqref{2-Gjk}. It also coincides, at
$\beta=0$, with the $N$-particle density of the ground state
$\hat\rho(\lambda)$. It is easy to check that both functions
$\hat\rho_\kappa(\lambda)$ (for $\kappa\ne 1$) and
$\hat\rho(\lambda)$ go to the same function $\rho(\lambda)$
\eqref{rho-lim}  in the thermodynamic limit.

Using the counting functions for the ground state \eqref{BE-xi} and twisted ground state \eqref{TBE-cf}, one can define shift functions 
describing the displacement of the spectral parameters $\{\mu\}$ with respect to ground state ones $\{\lambda\}$.
 One can actually introduce two shift functions:
 \begin{equation}\label{SP-shiftF}
 \hat F(\lambda_j)=M\bigl[\hat\xi(\mu_j)-\hat\xi(\lambda_j)\bigr], \qquad
 \hat F_\kappa(\mu_j)=M\bigl[\hat\xi_\kappa(\mu_j)-\hat\xi_\kappa(\lambda_j)\bigr].
 \end{equation}
Both have the same thermodynamic limit which we, from now on, denote by $F(\lambda)$.
One can easily verify that
 \begin{equation}\label{Shift-lim}
 F(\lambda_j)=\lim_{N,M\to\infty}\frac{\mu_{j}-\lambda_j}{\lambda_{j+1}-\lambda_j}.
 \end{equation}
The subtraction of the equation \eqref{BE}  with $n_j=I_j$ (Bethe
equation for the ground state) from \eqref{TBE} (Bethe equation for
the twisted ground state, corresponding to the same number
$\tilde{n}_j=I_j)$ and a replacement of finite differences by derivatives
and of sums by integrals as in \eqref{sum-int}, leads to the
integral equation for the shift function $F$:
 \begin{equation}\label{0-DC-beta}
 F(\lambda)+\frac1{2\pi}\int\limits_{-q}^qK(\lambda-\mu)F(\mu)\,\dd\mu
 =\frac{\beta}{2\pi i}.
 \end{equation}
Comparing this equation with the one for the dressed charge \eqref{0-DC}, one obtains
 \begin{equation}\label{Sh-DC}
 F(\lambda)=\frac{\beta Z(\lambda)}{2\pi i}.
 \end{equation}
Thence, for $M$ large enough, the spectral parameters $\mu_j$
are shifted with respect to the ground state ones $\lambda_j$ by a quantity $\eps_j=\mu_j-\la_j$
which can be computed for large $M$
 \begin{equation}\label{shift}
 \eps_j=\frac{\beta Z(\lambda_j)}{2\pi iM\rho(\lambda_j)}+O(M^{-2}),
 \qquad j=1,\dots,N.
 \end{equation}

\subsection{Thermodynamic limit of $ {\cal A}_N^{(\kappa)}$ and $C_{0,N}^{(\kappa)}$}
\label{sec-A-C}

We first compute the limits of $C_{0,N}^{(\kappa)}$ and ${\cal A}_N^{(\kappa)}$.
These limits are quite simple to calculate since it is enough to take into account only the
leading order of the shift \eqref{shift}.

We have already seen that, in the thermodynamic limit $N,M\to\infty$,
discreet sums over the parameters $\lambda_j$ can be replaced by
integrals $via$ \eqref{sum-int}. It is also easy to see that
 \begin{equation}\label{lim-prod-f}
 \lim_{N,M\to\infty}
   \prod_{j=1}^N\left(1+\frac1M  f(\lambda_j)\right)
   =\exp\left\{\int\limits_{-q}^q f(\lambda) \, \rho(\lambda)\,\dd\lambda\right\}  ,
 \end{equation}
\vspace{-5mm}
 \begin{equation}\label{lim-det-V}
 \lim_{N,M\to\infty}
   \det_N\left[\delta_{jk}+\frac1M V(\lambda_j,\lambda_k)\right]
   =\det[I+V\rho] \, ,
 \end{equation}
provided $f(\lambda)$ and $V(\lambda,\mu)$ are piecewise continuous on $[-q,q]$.
In \eqref{lim-det-V}, $\det[I+V\rho]$ refers to the Fredholm determinant
of the integral operator $I+V\rho$ acting on the interval $[-q,q]$ with kernel $V(\lambda,\mu)\rho(\mu)$.

The thermodynamic limit of $C_{0,N}^{(\kappa)}$ is straightforward to obtain using \eqref{shift}:
 \begin{align}
 \lim_{N,M\to\infty} C_{0,N}^{(\kappa)}
 &=\lim_{N,M\to\infty}\sum_{j,k=1}^N
 \log\frac{\sinh(\lambda_j-\lambda_k-\epsilon_k-i\zeta) \,
                 \sinh(\lambda_j-\lambda_k+\epsilon_j-i\zeta)}
                {\sinh(\lambda_j-\lambda_k-i\zeta) \,
                 \sinh(\lambda_j-\lambda_k+\epsilon_j-\epsilon_k-i\zeta)}
                  \nonumber\\
& =
 -\lim_{N,M\to\infty}\sum_{j,k=1}^N
   \epsilon_j \, \epsilon_k \;
   \partial_{\lambda_j} \partial_{\lambda_k}
   \log\sinh(\lambda_j-\lambda_k-i\zeta)
 =\frac{\beta^2}{4\pi^2}C_0,
 \label{lim-W}
 \end{align}
where $C_0$ is given by \eqref{7-C0}.

Let us now consider the different factors in \eqref{6-ev-phiC}.
As the limit of  $\hat\rho(\lambda)$ and
$\hat\rho_\kappa(\lambda)$ coincides with the ground state
density $\rho(\lambda)$, both
determinants $\det_N\Theta_{jk}^{(\lambda,\mu)}$ go to the Fredholm
determinant of the operator $I+\tf{K}{2\pi}$:
 \begin{equation}\label{lim-norm}
 \lim_{N,M\to\infty}\det_N\Theta_{jk}^{(\lambda,\mu)}=
 \det\left[I+\tf{K}{2\pi} \right] .
 \end{equation}
A direct application of \eqref{lim-prod-f} leads to
\begin{equation}\label{lim-Vpm}
 \lim_{N,M\to\infty}V_{\pm}(w)=  e^{-\beta\tilde z(w\pm i\zeta)}.
\end{equation}
Similarly,
 \begin{equation}\label{lim-V0}
 \lim_{N,M\to\infty}\prod_{a=1}^{N}
 \frac {\sinh(w-\lambda_a)}{\sinh(w-\mu_a)}=e^{-\beta\tilde z(w)},
 \qquad w \, \text{uniformly away from} \; [-q,q].
 \end{equation}
Substituting \eqref{lim-Vpm} and  \eqref{lim-V0} into \eqref{2-U1} we have that, at $\theta_1=-q$,
 \begin{equation}\label{lim-kern1}
 \lim_{N,M\to\infty}\frac1{2\pi i}\widehat U_{-q}^{(\lambda)}(w,w')
 =\frac1{2\pi i}U_{-q}^{(\lambda)}(w,w'),
 \end{equation}
where the kernel $U_{-q}^{(\lambda)}(w,w')$ is given by
\eqref{7-U1}. It is also easy to see that, when $\theta_2=q$ and
$\beta$ is purely imaginary,
 \begin{equation}\label{lim-kern2}
 \lim_{N,M\to\infty}\frac1{2\pi i}\widehat U_{q}^{(\mu)}(w,w')
 =  \left(\frac1{2\pi i}U_{-q}^{(\lambda)}\right)^\dagger(-\bar w,-\bar w'),
\end{equation}
where $\dagger$ means hermitian conjugation. A simple algebra based on
\eqref{6-prop-z1} leads to
 \begin{equation}\label{lim-A}
 \lim_{N,M\to\infty}{\cal A}_N^{(\kappa)}=\left(\frac{\sinh\frac{\beta}2}{\sinh\frac{\beta{\cal Z}}2}\right)^2\widetilde{\cal
 A}(\beta),
 \end{equation}
where $\widetilde{\cal A}(\beta)$ is given by \eqref{7-const-A}.

\subsection{Thermodynamic limit of the Cauchy determinant}
\label{sec-D}

As already mentioned, the thermodynamic limit of the factor
$D_N^{(\kappa)}$ \eqref{6-DNk} is the most complicated to obtain.
Although the
final result for this quantity can be expressed in terms of the
limiting value of the shift function \eqref{Sh-DC}, the
approximation \eqref{shift} is not enough for the intermediate analysis. The reason is that
$D_N^{(\kappa)}$ contains the square of a Cauchy determinant
which has poles at $\mu_j=\lambda_k$.
These singularities are compensated by the zeros of
 \begin{equation}\label{7-prod}
 \prod\limits_{j=1}^N
 \left(\frac{\kappa V_+(\lambda_j)}{V_-(\lambda_j)}-1\right)
 \left(\frac{V_-(\mu_j)}{\kappa V_+(\mu_j)}-1\right) \; .
 \end{equation}
However, this compensation takes place if and only if the parameters
$\mu_j$  {\em exactly} solve the set of twisted Bethe equations,
which means that  it is not enough to consider only the first order of the shift function as in \eqref{shift}. Therefore, to compute the limit of
$D_N^{(\kappa)}$, we should work with the exact formulae for the $N$-particle shift
functions $\hat F$ and $\hat F_\kappa$ \eqref{SP-shiftF}
defined in terms of the two counting functions
$\hat\xi(\omega)$ and $\hat\xi_\kappa(\omega)$ \eqref{BE-xi}, \eqref{TBE-cf},
their
limiting values being only taken at the end of the computation.

In order to re-write \eqref{6-DNk} in terms of the shift functions $\hat F$ and $\hat F_\kappa$,
we also introduce the functions
 \begin{equation}\label{XXZ-Uniform}
 \hat\varphi(\lambda,\mu)=\frac{\sinh(\lambda-\mu)}{\hat\xi(\lambda)-\hat\xi(\mu)},
 \qquad
 \hat\varphi_\kappa(\lambda,\mu)=\frac{\sinh(\lambda-\mu)}{\hat\xi_\kappa(\lambda)-\hat\xi_\kappa(\mu)},
 \end{equation}
which fulfill
 \begin{equation}\label{SP-pi-rho}
 \hat\varphi^{-1}(\lambda,\lambda)=\hat\xi'(\lambda)=\hat\rho(\lambda),
 \qquad
 \hat\varphi^{-1}_\kappa(\lambda,\lambda)=\hat\xi'_\kappa(\lambda)=\hat\rho_\kappa(\lambda).
 \end{equation}
Note that $\hat\varphi$ and $ \hat\varphi_{\kappa}$ have the same thermodynamic limit denoted by $\varphi$.

\begin{prop}\label{P-factor}
The product $D_N^{(\kappa)}$ \eqref{6-DNk} admits the
following factorization:
 \begin{equation}\label{7-DNk}
 D_N^{(\kappa)}
 =H[\hat F] \cdot H[-\hat F_\kappa] \cdot
   \Phi[\hat\varphi] \cdot \Phi[\hat\varphi_\kappa] \cdot
   \prod_{j=1}^N\frac{\hat\varphi(\lambda_j,\lambda_j) \, \hat\varphi_\kappa(\mu_j,\mu_j)}
                                   {\hat\varphi(\lambda_j,\mu_j) \, \hat\varphi_\kappa(\mu_j,\lambda_j)}.
 \end{equation}
The functions $H[\hat F]$ and $ \Phi[\hat\varphi]$ are defined as
 \begin{equation}\label{7-HF}
 H[\hat F]=
 \prod\limits_{j>k}^N \left(1+\frac{\hat F(\lambda_j)-\hat F(\lambda_k)}{j-k}\right)
 \prod\limits_{j,k=1\atop{j\ne
 k}}^N\left(1+\frac{\hat F(\lambda_j)}{j-k}\right)^{-1}\prod_{j=1}^N\frac{\sin\bigl(\pi
 \hat F(\lambda_j)\bigr)}{\pi
 \hat F(\lambda_j)},
 \end{equation}
\vspace{-5mm}
 \begin{equation}\label{Fi-Fi}
 \Phi[\hat\varphi]
 =\frac{\prod\limits_{j>k}^N \big[ \hat\varphi(\lambda_{j},\lambda_{k}) \,
                                                \hat\varphi(\mu_{k},\mu_j) \big] }
    {\prod\limits_{j,k=1\atop{j\ne k}}^N\hat\varphi(\mu_j,\lambda_{k})},
 \end{equation}
and the function $H[-\hat F_\kappa]$ (respectively $\Phi[\hat\varphi_\kappa]$) can be obtained from \eqref{7-HF} (resp.  \eqref{Fi-Fi}) by
the replacement $\hat F(\lambda)\to -\hat F_\kappa(\mu)$
(resp.  $\hat\varphi\to\hat\varphi_\kappa$).
\end{prop}

\Proof
Using the definition \eqref{BE-xi}, \eqref{TBE-cf} of the counting
functions, one can easily see that
 \begin{equation}\label{V+-xi}
 \frac{\kappa V_+(\omega)}{V_-(\omega)}=e^{2\pi
 iM [\hat\xi(\omega)-\hat\xi_\kappa(\omega)]},
 \end{equation}
which leads to
 \begin{align}
 &\prod_{j=1}^N\left(\frac{\kappa V_+(\lambda_j)}{V_-(\lambda_j)}-1\right)
 \left(\frac{V_-(\mu_j)}{\kappa V_+(\mu_j)}-1\right)
      \nonumber\\
 &\hspace{3cm}
   = \prod_{j=1}^N\left(e^{2\pi
 iM [\hat\xi(\lambda_j)-\hat\xi_\kappa(\lambda_j) ]}-1\right)
 \left(e^{-2\pi
 iM [\hat\xi(\mu_j)-\hat\xi_\kappa(\mu_j) ]}-1\right)
        \nonumber\\
&\hspace{3cm}
  = \prod_{j=1}^N\left(e^{2\pi
 iM [\hat\xi_\kappa(\mu_j)-\hat\xi_\kappa(\lambda_j) ]}-1\right)
 \left(e^{-2\pi
 iM [\hat\xi(\mu_j)-\hat\xi(\lambda_j) ]}-1\right)
        \nonumber\\
 &\hspace{3cm}
   =\prod_{j=1}^N\left[
 4\sin\bigl(\pi \hat F_\kappa(\mu_j)\bigr)\sin\bigl(\pi
 \hat F(\lambda_j)\bigr)\right].
 \label{prod-prod}
 \end{align}
Here we have used that $\hat\xi(\lambda_j)=\hat\xi_\kappa(\mu_j)=j/M$,
 and the identity
 \begin{equation}\label{SP-ident}
 \sum_{j=1}^N\Bigl[\hat\xi(\lambda_j)-\hat\xi(\mu_j)+\hat\xi_\kappa(\mu_j)-\hat\xi_\kappa(\lambda_j)\Bigr]=0
 \end{equation}
which is a consequence of  \eqref{BE-xi}, \eqref{TBE-cf}.

The remaining part of the proof is quite trivial. It is based on the expression for the Cauchy determinant in terms of double
products
 \begin{equation}\label{7-Cauchy}
 \det_N\left(\frac1{u_j-v_k}\right)=\frac{\prod\limits_{j>k}^N (u_j-u_k)
 (v_{k}-v_j)}{\prod\limits_{j,k=1}^N(u_j-v_k)},
 \end{equation}
where $u_j$, $v_j$ are arbitrary complex numbers.
Substituting  the explicit expressions for all the
factors into \eqref{7-DNk}, and using again
$\hat\xi(\lambda_j)=\hat\xi_\kappa(\mu_j)=j/M$ together with the
definition of $\hat F$ and $\hat F_\kappa$ \eqref{SP-shiftF}, one
easily reproduces the original representation \eqref{6-DNk} for
$D_N^{(\kappa)}$.
\qed

Since the functions $\hat\varphi(\lambda,\mu)$ and
$\hat\varphi_\kappa(\lambda,\mu)$ are not singular at
$\lambda,\mu\in[-q,q]$, the thermodynamic limits of the last three
factors in the r.h.s. of \eqref{7-DNk}  can easily be calculated similarly as in the previous subsection.
Using \eqref{shift} we find
 \begin{multline}\label{Td-Phi}
 \lim_{N,M\to\infty}\Phi[\hat\varphi] \cdot \Phi[\hat\varphi_\kappa] \cdot
 \prod_{j=1}^N\frac{\hat\varphi(\lambda_j,\lambda_j) \, \hat\varphi_\kappa(\mu_j,\mu_j)}
                                 {\hat\varphi(\lambda_j,\mu_j) \, \hat\varphi_\kappa(\mu_j,\lambda_j)}
                                 \\
  =
 \exp\Bigg\{-\frac{\beta^2}{4\pi^2}\int\limits_{-q}^q
    Z(\lambda)\, Z(\mu)\;
    \partial_\lambda \partial_\mu \log\varphi(\lambda,\mu) \;
    \dd\lambda\, \dd\mu   \Bigg\}.
 \end{multline}

It remains to compute the limiting values of $H[\hat F]$, $H[-\hat
F_{\kappa}]$. To do this, it is convenient to split $H[\hat F]$
into two factors: $H[\hat F]=H_1[\hat F]\cdot H_2[\hat F]$, with
 \begin{equation}\label{7-H1}
 H_1[\hat F]
 =\exp\Bigg[- \sum_{j=1}^N\psi(j)\bigl[\hat F(\lambda_j)-\hat F(\lambda_{N-j+1})\bigr]\Bigg]
 \cdot\prod\limits_{j>k}^N
 \left(1+\frac{\hat F(\lambda_j)-\hat F(\lambda_k)}{j-k}\right),
 \end{equation}
\vspace{-5mm}
 \begin{equation}\label{7-H2}
 H_2[\hat F]
 =\exp\Bigg[ \sum_{j=1}^N\psi(j)\bigl[\hat F(\lambda_j)-\hat F(\lambda_{N-j+1})\bigr]\Bigg]
 \prod\limits_{j,k=1\atop{j\ne k}}^N \!\! \left(1+\frac{\hat F(\lambda_j)}{j-k}\right)^{\! -1} \!
 \prod_{j=1}^N\frac{\sin[\pi \hat F(\lambda_j)]}{\pi \hat F(\lambda_j)}, \hspace{-1mm}
 \end{equation}
$\psi(z)$ being the logarithmic derivative of the
$\Gamma$-function (\textit{cf} Appendix \ref{Appendix The psi function }).

\begin{prop}\label{P-H1}
The thermodynamic limit of the factor $H_1[\hat F]$ \eqref{7-H1} is given by
 \begin{equation}\label{lim-H1}
 \lim_{N,M\to\infty} \log H_1[\hat F]=-\frac14
 \int\limits_{-q}^q\left(\frac{F(\lambda)-F(\mu)}{\xi(\lambda)-\xi(\mu)}\right)^2
 \rho(\lambda)\,\rho(\mu)\; \dd\lambda\,\dd\mu.
 \end{equation}
\end{prop}

\Proof
Due to the smoothness of $F$ and $\xi$, there is an $\eps$, $0<\eps<1$, such that, for $M$ large enough,
 \begin{equation}\label{Est}
 \left|\frac{\hat F(\lambda_j)-\hat F(\lambda_k)}{j-k}\right|=
 \frac1M\left|\frac{\hat F(\lambda_j)-\hat F(\lambda_k)}{\hat\xi(\lambda_j)-\hat\xi(\lambda_k)}\right|<\eps \; ,\quad 
\e{uniformy} \, \e{in} \, \la_j , \, \la_k .
 \end{equation}
The logarithm of the double product in the r.h.s. of
\eqref{7-H1} can therefore be expanded into its Taylor integral series of order 2
 \begin{multline}\label{log1}
 \log \prod\limits_{j>k}^N
 \left(1+\frac{\hat F(\lambda_j)-\hat F(\lambda_k)}{j-k}\right)
 =
 \sum_{n=1}^{2}\frac{(-1)^{n+1}}{nM^n}\sum_{j>k}^N
 \left(\frac{\hat F(\lambda_j)-\hat F(\lambda_k)}
                {\hat\xi(\lambda_j)-\hat\xi(\lambda_k)}\right)^n \\
+ \frac{1}{M^{3}} \sum_{j>k}^N \Int{0}{1} \dd t
   \frac{\pa{1-t}^2}{ \Big(1+\f{t}{M}\cdot
         \frac{\hat F(\lambda_j)-\hat F(\lambda_k)}
                {\hat\xi(\lambda_j)-\hat\xi(\lambda_k)}  \Big)^3 }
 \left(\frac{\hat F(\lambda_j)-\hat F(\lambda_k)}{\hat\xi(\lambda_j)-\hat\xi(\lambda_k)}\right)^3.
 \end{multline}
The expressions under the double sum over $j$ and $k$ being not
singular at $\lambda_j,\lambda_k\in [-q,q]$, the
thermodynamic limit of the integral remaining terms in \eqref{log1}
vanish due to the $M^{-3}$ pre-factor. The contribution issued from the $n=1$ term
cancels with the pre-factor in \eqref{7-H1} since
 \begin{equation}\label{Trans-n=1}
 \sum_{j>k}^N\frac{\hat F(\lambda_j)-\hat F(\lambda_k)}{j-k}=
 \sum_{j,k=1 \atop{ j\ne k}}^N\frac{\hat F(\lambda_j)}{j-k}
 =\sum_{j=1}^N\psi(j)\bigl[\hat F(\lambda_j)-\hat F(\lambda_{N-j+1})\bigr].
 \end{equation}
Thus, the limit of $\log H_1$ reduces to the one of the term $n=2$
in the expansion \eqref{log1}, which ends the proof.
\qed

\begin{prop}\label{P-H2}
The thermodynamic limit of the factor $H_2 [\hat F]$  \eqref{7-H2} is given by
 \begin{multline}\label{7-limH2}
 \lim_{N,M\to\infty}
 \left(H_2[\hat F] \cdot N^{\frac{F^2(-q)+F^2(q)}2}\right)
 =
 G\bigl(1+F(-q)\bigr) \, G\bigl(1-F(q)\bigr) \cdot e^{\frac{F(-q)-F(q)}2(1-\log
 2\pi)}\\
 \times
 \exp\left\{\int\limits_{-q}^q\left[\frac{F^2(\lambda)-F^2(q)}{\xi(\lambda)-\xi(q)}
 -\frac{F^2(\lambda)-F^2(-q)}{\xi(\lambda)-\xi(-q)}\right]
 \rho(\lambda)\, \dd\lambda\right\} ,
 \end{multline}
where $G(z)$ is the Barnes function \eqref{Barnes-A}.
\end{prop}

\Proof
The product over $k$ in \eqref{7-H2} can be expressed in terms of $\Ga$-functions as
 \begin{equation}\label{2-prod}
  \prod_{j,k=1\atop{j\ne  k}}^N
 \left(1+\frac{\hat F(\lambda_j)}{j-k}\right)^{-1}
 \prod_{j=1}^N\frac{\sin(\pi \hat F(\lambda_j))}{\pi \hat F(\lambda_j)}=
 \prod_{j=1}^N\frac{\Gamma^2(j)}{\Gamma\bigl(j+\hat F(\lambda_j)\bigr)\Gamma\bigl(j-\hat F(\lambda_{N-j+1})\bigr)}.
 \end{equation}
This means that $H_2 [\hat F]$ can itself be factorized into two factors,
 \begin{equation}\label{H-WW}
 H_2[\hat F]= H_2^{(1)} [\hat F] \cdot H_2^{(2)} [\hat F ],
 \end{equation}
with
 \begin{equation}\label{7-W}
H_2^{(1)}[\hat F] =\prod_{j=1}^N\frac{\Gamma(j)}{\Gamma(j+\hat F(\lambda_j))}\,
                                                          e^{\psi(j)\,\hat F(\lambda_j)}  ,
 \end{equation}
$H_2^{(2)}[\hat F]$ being obtained from \eqref{7-W} by the replacement $\hat F(\la_{j}) \tend - \hat F(\la_{N-j+1})$.

We first consider $H_2^{(1)}[\hat F]$. Let  $\hat F_-=\hat
F\bigl(\hat\xi^{-1}(0)\bigr)$. Evidently, $\hat F_-\to F(-q)$ in the
thermodynamic limit.  We use the fact that the shift function $\hat F(\lambda_j)$ is bounded uniformly for any value of $N$.  
Thus we can use the following bound $|\hat F(\lambda_j)|<n$ for some
$n\in\mathbb{N}$ and $j=1,\dots,N$. Then we represent
$H_2^{(1)}[\hat F]$ as
 \begin{equation}\label{splitting}
 H_2^{(1)}[\hat F]
 = \prod_{j=1}^N\frac{\Gamma(j)e^{\psi(j)\hat F_-}}{\Gamma(j+\hat F_-)}
 \prod_{j=1}^{n-1}\frac{\Gamma(j+\hat F_-)e^{( \hat F(\lambda_j)-\hat F_-)\psi(j)}}{\Gamma\bigl(j+\hat F(\lambda_j)\bigr)}
 \prod_{j=n}^{N}\frac{\Gamma(j+\hat F_-)e^{( \hat F(\lambda_j)-\hat F_-)\psi(j)}}{\Gamma\bigl(j+\hat
 F(\lambda_j)\bigr)}.
 \end{equation}
Using \eqref{simple-sumG}, \eq{a00} we find
  \begin{equation}\label{1-st-prod}
 \prod_{j=1}^N\frac{\Gamma(j)e^{\psi(j)\hat F_-}}{\Gamma(j+\hat F_-)}=
 \frac{G(N+1) \, G (1+\hat F_-)}{G(N+1+\hat F_-)}e^{\hat
 F_-(N\psi(N)+1-N)},
   \end{equation}
and thus, applying the asymptotic formulae \eqref{psi-asy},
\eqref{asy-Barns} we arrive at
 \begin{equation}\label{lim-1-st-prod}
 \lim_{N,M\to\infty}N^{\frac{\hat F_-^2}2}\prod_{j=1}^N\frac{\Gamma(j)e^{\psi(j)\hat F_-}}{\Gamma(j+\hat F_-)}=
 G\bigl(1+F(-q)\bigr)\cdot e^{ \frac{F(-q)}2(1-\log 2\pi)}.
 \end{equation}

The limit of the second finite product in \eqref{splitting} evidently is
equal to $1$ due to $\hat F(\lambda_j)-\hat F_-=O(M^{-1})$ for
$j=1,\dots,n-1$. Finally, in order to compute the limit of the last
product in \eqref{splitting} we can use Taylor integral representation for the
logarithms of $\Gamma$-functions (recall that $|\hat
F(\lambda_j)|<n\le j$)

 \begin{multline}
 \prod_{j=n}^{N}\frac{\Gamma(j+\hat F_-)e^{( \hat F(\lambda_j)-\hat F_-)\psi(j)}}{\Gamma\bigl(j+\hat
 F(\lambda_j)\bigr)}
 =\exp\left\{ \frac12\sum_{j=n}^{N}\psi'(j)
 \pac{\hat F^{2}_--\hat F^{2}(\la_j)}\right.\\
 \left.- \frac12\sum_{j=n}^{N}\Int{0}{1}
 \dd t (1-t)^{2} \pac{ \psi''\big(
 \hat F(\la_j)t+j)\, \hat F^{3}(\la_j) -\psi''\big(\hat F_- t
 +j\big)\, \hat F^{3}_- }  \right\}. \label{H-mod}
 \end{multline}
The limit of the first sum  in \eqref{H-mod} is obtained by applying
Lemma~\ref{m>1} proved in Appendix \ref{Ssf}. Indeed, as the
counting function is invertible and satisfies
$\hat\xi(\lambda_j)=j/M$, we write $\hat F(\lambda_j)= \hat
F\big(\hat\xi^{-1}(\frac jM)\big)$. Then, using \eqref{Lim-Gm}, we
find
 \begin{equation}\label{est-3}
 \lim_{N,M\to\infty}\sum_{j=n}^N\psi'(j)\,
         \bigl[\hat F_-^2-\hat F^2(\lambda_j)\bigr]
 = - \int_0^D\frac{F^2\bigl(\xi^{-1}(x)\bigr)-F^2\bigl(\xi^{-1}(0)\bigr)}x\, \dd x.
 \end{equation}
Recall that $D=\lim_{N,M\to\infty}N/M$ is the average density, so
that, after the change of variables $\xi^{-1}(x)=\lambda$, we obtain
 \begin{equation}\label{est-4}
 \lim_{N,M\to\infty}\sum_{j=1}^N\psi'(j)\, \bigl[\hat F_-^2-\hat F^2(\lambda_j)\bigr]
 = - \int\limits_{-q}^q\frac{F^2(\lambda)-F^2(-q)}{\xi(\lambda)-\xi(-q)}\rho(\lambda)\, \dd\lambda  .
 \end{equation}

Finally, the same transformation on the function $F$ allows us to
apply, uniformly in $t$, Lemma \ref{reminder} to the second sum in \eqref{H-mod}, hence
proving that its thermodynamic limit is zero.

Substituting these results into \eqref{splitting}, we obtain that
 \begin{multline}\label{W-fin1}
 \lim_{N,M\to\infty}\left(H_2^{(1)} [\hat F] \cdot N^{\frac{F^2(-q)}2}\right)\\
 =G\bigl(1+F(-q)\bigr)\cdot
 \exp\Bigg[\frac{F(-q)}2(1-\log 2\pi)
 -
 \frac12\int\limits_{-q}^q\frac{F^2(\lambda)-F^2(-q)}{\xi(\lambda)-\xi(-q)}\,
             \rho(\lambda)\,\dd\lambda \Bigg].
 \end{multline}

The thermodynamic limit of $H_2^{(2)}[\hat F]$ can be
deduced from the previous calculation due to the fact that
$\hat F(\lambda_{N-j+1})=\hat F\big(\hat\xi^{-1}(\frac{N+1}M-\frac jM)\big)$.
\qed

The  limit of $H[-\hat F_\kappa]$  is obtained in the same spirit as above.
Since the limits of $\hat F$, $\hat\rho$ and
$\hat\xi$ coincide with those of their $\kappa$-deformed
analogs, the result is given by
\eqref{lim-H1} and \eqref{7-limH2} provided  that one makes the replacement
$F\to -F$.

This enables us to obtain the thermodynamic limit of  $D_N^{(\kappa)}$:

\begin{prop}
The product $D_N^{(\kappa)}$ admits the following thermodynamic limit:
 \begin{equation}\label{7-limD-res}
 \lim_{N,M\to\infty}\left(D_N^{(\kappa)} \; M^{-\frac{{\beta^2\cal Z}^2}{2\pi^2}}\right)
 =
 G^2\big(1,{\textstyle\frac{\beta{\cal Z}}{2\pi i}}\big)
 \,
 [\rho(q)\,\sinh2q]^{\frac{{\beta^2\cal Z}^2}{2\pi^2}}
 \,
 e^{-\frac{\beta^2}{4\pi^2}C_1},
 \end{equation}
where $C_1$ is given by \eqref{7-C_1}.
 We remind that $G(a,z)=G(a-z)G(a+z)$ and ${\cal Z}=Z(q)$.
 \end{prop}

\Proof
Using the expression  \eqref{Sh-DC} of $F$ and the symmetry properties of $Z, \rho$ and
$\xi$,
\begin{equation*}
  Z(\lambda)=Z(-\lambda),\quad \rho(\lambda)=\rho(-\lambda),\quad
 \text{and}\quad \xi(\lambda)=D-\xi(-\lambda),
\end{equation*}
we obtain from Propositions~\ref{P-H1}, \ref{P-H2} that
 \begin{multline}\label{7-limD}
 \lim_{N,M\to\infty}\left(D_N^{(\kappa)} \; N^{-\frac{\beta^2{\cal Z}^2}{2\pi^2}}\right)
 =
 G^2\big(1,{\textstyle\frac{\beta{\cal Z}}{2\pi i}}\big)\;
 \exp   \Bigg[\frac{\beta^2}{8\pi^2}
                 \int\limits_{-q}^q
                 \left(\frac{Z(\lambda)-Z(\mu)}{\xi(\lambda)-\xi(\mu)}\right)^2
                 \rho(\lambda)\,\rho(\mu)\, \dd\lambda\, \dd\mu \Bigg]\\
 \times
 \exp\Bigg[ -\frac{\beta^2}{4\pi^2}\int\limits_{-q}^q
    Z(\lambda) Z(\mu)\,
    \partial_\lambda \partial_\mu\log\varphi(\lambda,\mu)
 \, \dd\lambda\,\dd\mu
 -\frac{\beta^2}{2\pi^2}\int\limits_{-q}^q
    \frac{Z^2(\lambda)-{\cal Z}^2}{\xi(\lambda)-\xi(q)}
    \rho(\lambda)\, \dd\lambda\Bigg] .
 \end{multline}
After some simple algebra the combination of integrals in
\eqref{7-limD} can be reduced to the constant $C_1$ \eqref{7-C_1},
and the result takes the announced form. \qed

\subsection{Results for the normalized scalar product and special form factor}

Putting together the individual limits \eqref{7-limD-res}, \eqref{lim-W} and
\eqref{lim-A} yields
 \begin{equation}\label{lim-S}
 \lim_{N,M\to\infty}\left(S_N^{(\kappa)}\cdot M^{-\frac{{\beta^2\cal Z}^2}{2\pi^2}}\right)
 =
 \left[\frac{2\sinh\frac{\beta} 2}{\beta{\cal Z}}\;
        G\big(2,{\textstyle\frac{\beta{\cal Z}}{2\pi i}}\big)\right]^2
                       \big[\rho(q)\, \sinh2q \big]^{\frac{{\beta^2\cal Z}^2}{2\pi^2}} \,
 e^{-\frac{\beta^2}{4\pi^2}(C_1-C_0)} \, \widetilde{\cal A}(\beta).
 \end{equation}
The limits being uniform in $\be$ and its derivatives, we take the
second $\be$ derivative at $\be=2\pi i$ and obtain
 \begin{equation}\label{lim-S-der}
 \lim_{N,M\to\infty}
 \left(-\left.\partial^2_\beta S_N^{(\kappa)}\right|_{\beta=2\pi i}\cdot M^{2{\cal Z}^2}\right)
 =
 \frac12\left(\frac{G(2,{\cal Z})}{\pi{\cal Z}}\right)^2
          \frac{ e^{C_1-C_0} \,
 \Bigl.\widetilde{\cal A}(\beta)}{ \big[ \rho(q)\, \sinh2q\big]^{2{\cal Z}^2}}\Bigr|_{\beta=2\pi i}.
 \end{equation}

We still need  to compute the excitation
momentum ${\cal P}_{ex}$ in \eqref{m-el-sigma} to finish our proof. We have
 \begin{equation}\label{lim-P}
 \lim_{N,M\to\infty}{\cal P}_{ex}
 =\lim_{N,M\to\infty}\sum_{j=1}^N \bigl[p_0(\mu_j)-p_0(\lambda_j)\bigr]
 =\int\limits_{-q}^q p'_0(\lambda) \, Z(\lambda)\, \dd\lambda.
 \end{equation}
The integral equations \eqref{0-rho}, \eqref{0-DC} for the density
and the dressed charge then lead to
 \begin{equation}\label{lim-P1}
 \int\limits_{-q}^q p'_0(\lambda) \, Z(\lambda)\, \dd\lambda
  = 2\pi  \int\limits_{-q}^q\rho(\lambda)\, \dd\lambda
  = 2p_{{}_F},
 \end{equation}
due to \eqref{0-Dmom}. Substituting \eqref{lim-P1} and \eqref{lim-S-der} into
\eqref{m-el-sigma} and comparing the result with the amplitude
\eqref{7-corr-funct-1}, we finally arrive at \eqref{m-el-sigmaA}.

\section*{Conclusion}

In the present article, we have initiated the development of a method to study form factors in the massless regime of the XXZ Heisenberg spin chain. In particular we have given explicitly the thermodynamic limit of
a special form factor in the XXZ chain. This form factor corresponds to the matrix
element of the operator $\sigma^z$ between the ground state and an
excited state containing one particle and one hole on the different
ends of the Fermi zone. It is clear however that our method 
can be applied to other form factors as well, where excited states
contain arbitrary number of particles and holes in arbitrary
positions. This method can also be applied to other integrable models,
for example to the system of one-dimensional bosons.

One of the unsolved problems is to prove that at zero magnetic field our
result coincides with the amplitude predicted in \cite{Luk99,LukT03}. The
limit $h\tend 0$ corresponds to the limit $q\tend\infty$ in
\eqref{lim-S}. It is easy to see that the constants $C_0$, $C_1$,
$\widetilde{\cal A}$, $\rho(q)\sinh2q$, being finite for finite $q$,
become divergent at $q\tend\infty$. We were able to prove that the
total combination giving $F_\sigma$ remains finite for $h=0$, but we
did not obtain a simple expression for its value in this limit. We succeeded nevertheless to
compute explicitly this quantity  in the vicinity of the free
fermion point ($\zeta=\frac\pi2$) up to the second order in
$\epsilon=\frac\zeta\pi-\frac12$. Our computation confirms the
result \cite{Luk99,LukT03} up to that order.

\section*{Acknowledgements}

J. M. M., N. S. and V. T. are supported by CNRS.
N. K., K. K. K., J. M. M. and V. T. are supported by the ANR program GIMP
ANR-05-BLAN-0029-01.
N. K. and V. T. are supported by the ANR program MIB-05 JC05-52749. We also acknowledge  the support from the GDRI-471 of CNRS "French-Russian network in Theoretical and Mathematical  Physics". N. S. is also supported by 
the Program of RAS Mathematical Methods of the Nonlinear Dynamics,
RFBR-08-01-00501a, Scientific Schools 795.2008.1.
K. K. K. is supported by the French ministry of research. 
N. K and N. S. would like to thank the Theoretical Physics group of the
Laboratory of Physics at ENS Lyon for hospitality, which makes this
collaboration possible.


\appendix

\section{$\psi$-function and Barnes function}

We recall in this appendix the definitions of the $\psi$-function
and the Barnes function, as well as several standard formulae that
are useful for our study.

\subsection{The $\psi$-function and its derivatives}
\label{Appendix The psi function }

The $\psi$-function is defined as the logarithmic derivative $\psi(z)=\f{\dd}{\dd z}\log\Gamma(z)$ of the $\Gamma$-function.
The multiplication property of the $\Gamma$-function implies that,
 \begin{equation}\label{diff-psi}
 \psi^{(n)}(z+1)-\psi^{(n)}(z)=\frac{(-1)^n n!}{z^{n+1}}.
 \end{equation}
It follows from \eqref{diff-psi} that
 \begin{equation}\label{simple-sum}
 \sum_{k=0}^{N-1}\frac1{(k+a)^{n+1}}=\frac{(-1)^n}{ n!}\Bigl(\psi^{(n)}(N+a)-\psi^{(n)}(a)\Bigr).
 \end{equation}
When $z\to\infty$ with $-\pi<\arg(z)<\pi$, one has
\begin{equation}\label{psi-asy}
 \psi(z)=\log z-\f{1}{2z}+ O\left(\frac{1}{z^2}\right),\qquad
 \psi^{(n)}(z)=\frac{(-1)^{n-1}(n-1)!}{z^n}+O\left(\frac1{z^{n+1}}\right).
 \end{equation}

For $n\geq 1$, the $n^{\rm{th}}$-derivative of the $\psi$-function admits the following integral representation:
 \begin{equation}\label{Int-rep}
 \psi^{(n)}(z)=-\int\limits_0^1\frac{x^{z-1}\log^nx}{1-x}\,\dd x.
 \end{equation}
The latter implies in particular that, for $z>0$,
 \begin{equation}\label{est-psi}
 (-1)^{n-1}\psi^{(n)}(z)>0, \qquad  n\ge 1.
 \end{equation}
It is also easy to see that
 \begin{equation}\label{est-psi-1z}
  \psi'(z)-\frac1z=\int\limits_0^1x^{z-1}\left(\frac{\log x}{x-1}-1\right)\,\dd x >0,
   \qquad
  z>0.
 \end{equation}
%

\subsection{The Barnes function}
\label{Appendix Barnes Function}

The Barnes function $G(z)$ is  the unique solution of
 \begin{equation}\label{Barnes-A}
 G(1+z)=\Gamma(z)\, G(z),
 \quad\mbox{with}\quad G(1)=1
 \quad\mbox{and}\quad
 \frac{\dd^3}{\dd z^3}\log G(z)\ge 0, \quad z>0 .
 \end{equation}
It has the following integral representation
 \begin{equation}\label{int-Barnes}
 \log G(1+z)=\frac{z(1-z)}2
 +\frac z2 \log2\pi
 +\int\limits_0^z x\, \psi(x)\, \dd x,
 \qquad \Re(z)>-1.
 \end{equation}
The Barnes function has the following asymptotic behavior when $z\to\infty$,
$-\pi<\arg(z)<\pi$:
 \begin{equation}\label{asy-Barns}
 \log G(1+z)=\left(\frac{z^2}2-\frac1{12}\right)\log z-
 \frac{3z^2}4+\frac z2\log
 2\pi 
 +\zeta'(-1)+O\Bigl(\frac1z\Bigr), 
 \end{equation}
Due to \eqref{Barnes-A}, one has
 \begin{equation}\label{simple-sumG}
 \sum_{k=0}^{N-1}\log\Gamma(k+a)=\log G(N+a)-\log G(a).
 \end{equation}
%
%
%

\section{Sums with logarithmic derivatives of $\Gamma$-function}
\label{Ssf}

In this appendix, we explain how to compute the thermodynamic limit
of some sums involving  the $\psi$-function and its derivatives.

\subsection{Finite sums involving $\psi$-function }

\begin{lemma}
For $n\ge 0$, $a\ne -1,-2,\dots$, and an arbitrary complex $\alpha$,
one has the following identity
 \begin{equation}\label{GF}
 \sum_{k=1}^N \psi^{(n)}(k+a)e^{\alpha k}
 =\frac1{e^\alpha-1}\cdot\Bigg[
 \psi^{(n)}(N+a)e^{\alpha (N+1)}- \psi^{(n)}(a)e^{\alpha} 
 -(-1)^n n!\sum_{k=1}^{N-1}\frac{e^{\alpha
 (k+1)}}{(k+a)^{n+1}}\Bigg].
 \end{equation}
\end{lemma}

\Proof  Denote the l.h.s. of \eqref{GF} as $f(\alpha)$,
 \begin{equation}\label{f-alpha}
 f(\alpha)=\sum_{k=1}^N \psi^{(n)}(k+a)\, e^{\alpha k}.
 \end{equation}
Shifting $k$ by $k+1$ and using \eqref{diff-psi}, we obtain
 \begin{align}
 f(\alpha) &=e^{\alpha}\sum_{k=0}^{N-1} \psi^{(n)}(k+a+1) \, e^{\alpha k}
         \nonumber\\
 &= e^{\alpha}\sum_{k=0}^{N-1} \psi^{(n)}(k+a)\, e^{\alpha k}
   +
 (-1)^n n!\sum_{k=0}^{N-1} \frac{e^{\alpha (k+1)}}{(k+a)^{n+1}}
        \nonumber\\
 &=e^{\alpha}f(\alpha)+e^{\alpha}\psi^{(n)}(a)-e^{\alpha (N+1)}\psi^{(n)}(N+a)+
 (-1)^n n!\sum_{k=0}^{N-1} \frac{e^{\alpha (k+1)}}{(k+a)^{n+1}}.
 \label{f-alpha1}
 \end{align}
Then \eqref{GF} follows immediately.  \qed

Taking derivatives of \eqref{GF} with respect to $\alpha$ at
$\alpha=0$ one obtains formulae for the sums of the type
$\psi^{(n)}(k+a)k^p$. Let us give explicitly several of them.

One has for $p=0$, $n=0$,
\begin{equation}\label{a00}
 \sum_{k=1}^N \psi(k+a)=(N+a)\psi(N+a)-a\psi(a)-N,
\end{equation}
and for $p=0$, $n\ge 1$,
 \begin{equation}
 \sum_{k=1}^N \psi^{(n)}(k+a)
   =(N+a)\psi^{(n)}(N+a)-a\psi^{(n)}(a+1)
    + n\Bigl(\psi^{(n-1)}(N+a)-\psi^{(n-1)}(a+1)\Bigr).\label{a0}
 \end{equation}
One also has for $p=1$, $n=1$,
 \begin{multline}
 \sum_{k=1}^N k\psi'(k+a)
   =\frac{(N+a)(N+1-a)}2\psi'(N+a)+\frac{a(a-1)}2\psi^{(n)}(a+1)\\
    -\frac{2a-1}2\Bigl(\psi(N+a)-\psi(a+1)\Bigr)
      +\frac{N-1}2,\label{a011}
 \end{multline}
and for $p=1$, $n\ge 2$,
 \begin{multline}
 \sum_{k=1}^N k\psi^{(n)}(k+a)
   =\frac{(N+a)(N+1-a)}2\psi^{(n)}(N+a)+\frac{a(a-1)}2\psi^{(n)}(a+1)\\
    -\frac{n(2a-1)}2\Bigl(\psi^{(n-1)}(N+a)-\psi^{(n-1)}(a+1)\Bigr)
      -\frac{n(n-1)}2\Bigl(\psi^{(n-2)}(N+a)-\psi^{(n-2)}(a+1)\Bigr).\label{a01}
 \end{multline}
%

\subsection{More complicated sums}

We now consider sums involving in addition some regular function(s).

\begin{lemma}\label{m>1}
 Let $f\in C^{1} ( [0,a] )$ for $a>0$.
 Let us consider the sum
 \begin{equation}
   S_1^{(N,M)}[f]
    =\sum_{k=1}^N\left[f\left({\textstyle\frac kM}\right)-f(0)\right]\psi'(k).
 \end{equation}
In the limit $N,M\tend \infty$, $N/M\tend D$ with $D\in[0,a[ $, it
tends to
 \begin{equation}\label{Lim-Gm}
S_1^{(N,M)}[f]  \tend  \Int{0}{D} \frac{ f(t)-f(0) }{t} \, \dd t .
 \end{equation}
\end{lemma}
\Proof We have
\begin{equation}\label{S1-new}
 S_1^{(N,M)}[f] = \sum_{k=1}^{N}\left(\psi'(k)-\frac1k\right)
 \frac kM \Int{0}{1} f'\!\paf{ k t
 }{M} \dd t + \sum_{k=1}^{N}
 \frac 1M \Int{0}{1} f'\!\paf{ k t
 }{M} \dd t.
 \end{equation}
It is easy to see that the first sum vanishes in the limit
considered. Indeed, setting  $\norm{f}=\sup_{[0,a] }\abs{f}$, we
get using \eqref{a011}, \eqref{est-psi-1z} and \eqref{psi-asy},
\begin{multline}\label{pouet}
\left| \sum_{k=1}^{N}\left(\psi'(k)-\frac1k\right)
 \frac kM \Int{0}{1} f'\!\paf{ k t
 }{M} \dd t\right|\leq \frac{\norm{f'}}M
 \sum_{k=1}^{N}\left(k\psi'(k)-1\right)\\
 =\frac{\norm{f'}}{2M}\left[N(N+1)\psi'(N)
 +\psi(N)-\psi(1)-N-1\right]=\frac{\norm{f'}}{2M}\left[
 \log N+O(1)\right].
 \end{multline}
Thus, the limit of $S_1^{(N,M)}[f]$ reduces to one of the second
term in \eqref{S1-new}, which is quite straightforward due to
Euler--Maclaurin summation formula
\begin{equation}
 \lim_{N,M\tend \infty}S_1^{(N,M)}[f] = \lim_{N,M\tend \infty}
 \sum_{k=1}^{N} \frac 1M \Int{0}{1} f'\!\paf{ k t }{M} \dd t
 = \Int{0}{D} \dd y\,\Int{0}{1}f'(yt) \dd t
 = \Int{0}{D} \f{ f\pa{y}-f\pa{0} }{ y } \dd y,
\end{equation}
and Lemma~\ref{m>1} is proved. \qed

\begin{lemma}\label{reminder}
Let $F,f \in C^{1} ([0,a] ) $, with $a>0$. Let
$n\in\mathbb{N}$ be such that $\norm{F}=\sup_{[0,a] }\abs{F}<n$.
Then, for $m>1$, the sum
\begin{equation}
\label{lim-Gm Sums}
  \wt{S}_{m}^{(N,M)}[f,F]
  = \sul{k=n}{N} \left[
        f\Big(\frac{k}{M}\Big)\, \psi^{\pa{m}}\Big(F\Big(\frac{k}{M}\Big) + k\Big)
         - f(0)\, \psi^{\pa{m}}\big(F(0)+k\big) \right]
\end{equation}
vanishes in the limit $N,M \tend \infty$, $N/M\tend D$ with $D\in
[0,a] $.
\end{lemma}

\Proof
We have
\begin{multline}
\wt{S}_{m}^{(N,M)}[f,F] =\sum_{k=n}^{N} \frac{k}{M}
\int\limits_{0}^{1} \dd t \Bigg[
     f'\Big(\frac{t k}{M}\Big)\, \psi^{\pa{m}}\Big(F\Big(\frac{t k}{M}\Big) + k\Big)
      \\
  + f\Big(\frac{t k}{M}\Big)\, F' \Big(\frac{t k}{M}\Big)\,
      \psi^{\pa{m+1}}\Big(F\Big(\frac{t k}{M}\Big) + k\Big)
      \Bigg].
\end{multline}
Due to \eqref{est-psi}, \eqref{est-psi-1z} $(-1)^{m-1}\psi^{(m)}(z)$
is positive monotonically decreasing function for $z>0$. Then one
has the following estimate
\begin{equation}
 |\wt{S}_{m}^{(N,M)}[f,F]|\le\f{\norm{F'}\norm{f}+ \norm{f'}}{M}
 \sul{k=n}{N} k\left[
  \pa{-1}^m \psi^{\pa{m+1}} (k-\norm{F})
   +
 \pa{-1}^{m-1} \psi^{\pa{m}} (k-\norm{F}) \right].
\end{equation}
The last sum is readily computed by \eqref{a01}. It is easy to see
that it is a $O(\log N)$ at most. \qed





\end{document}